\documentclass[]{jfm}
\usepackage{graphicx}
\usepackage{epstopdf, epsfig}
\usepackage[utf8]{inputenc} 
\usepackage[T1]{fontenc}   
\usepackage{bigints}
\usepackage{subfigure}
\usepackage{caption}
\usepackage{tikz}
\usepackage{color}

\newcommand{\blackline}{\raisebox{2pt}{\tikz{\draw[-,black,solid,line width = 1.5pt](0,0) -- (5mm,0);}}}

\newcommand{\greenline}{\raisebox{2pt}{\tikz{\draw[-,green,solid,line width = 1.5pt](0,0) -- (5mm,0);}}}
\newcommand{\greendashline}{\raisebox{2pt}{\tikz{\draw[-,green,dashed,line width = 1.5pt](0,0) -- (5mm,0);}}}
\newcommand{\magentaline}{\raisebox{2pt}{\tikz{\draw[-,magenta,solid,line width = 1.5pt](0,0) -- (5mm,0);}}}

\newcommand{\blackthinline}{\raisebox{2pt}{\tikz{\draw[-,black,line width = 0.5pt](0,0) -- (5mm,0);}}}

\shorttitle{Near-wall forces on a particle in a stagnation-point flow}
\shortauthor{J. Magnaudet \& M. Abbas}
\title{Near-wall forces on a neutrally-buoyant spherical particle in an axisymmetric stagnation-point  flow}
\author{Jacques Magnaudet\aff{1,2}
\corresp{\email{Jacques.Magnaudet@imft.fr}}
 \and Micheline Abbas\aff{3,2}
  \corresp{\email{Micheline.Abbas@ensiacet.fr}}}
\affiliation{\aff{1} Institut de M\'ecanique des Fluides de Toulouse (IMFT), Universit\'e de Toulouse, CNRS, Toulouse, France
\aff{2} FR FERMAT, Universit\'e de Toulouse, CNRS, Toulouse, France
\aff{3} Laboratoire de G\'enie Chimique (LGC), Universit\'e de Toulouse, CNRS, Toulouse, France}

\begin{document}

\maketitle
\begin{abstract}
Hydrodynamic forces acting on a neutrally-buoyant spherical particle immersed in a wall-bounded axisymmetric stagnation point flow (Hiemenz-Homann flow) are predicted, based on a suitable form of the reciprocal theorem. An approximate algebraic form of the undisturbed velocity field is set up, mimicking the gradual transition of the actual carrying flow throughout the boundary layer, from a pure linear straining flow in the bulk to a parabolic flow at the wall. The particle Reynolds number is assumed to be small and predictions based on the creeping-flow assumption are first derived. Then, inertial corrections are computed, assuming that the particle stands close enough to the wall for the latter to be in the inner region of the disturbance. Predictions for the time-dependent slip velocity between the particle and ambient fluid are obtained in the form of a differential equation, first assuming that the particle moves along the flow symmetry axis, then extending the analysis to particles released at an arbitrary radial position. 
In the former case, these predictions are compared with results provided by numerical simulations. When the strain-based Reynolds number (built on the particle radius and strain rate in the bulk) exceeds $0.1$, finite-inertia effects due to particle-wall interactions and to the relative acceleration between the particle and fluid are found to substantially modify the way the slip velocity varies with the distance to the wall. 
\end{abstract}

\section{Introduction}
After completing his monumental textbook on fluid dynamics, Batchelor turned his research into what he called micro-hydrodynamics, beginning a second scientific life. His most outstanding contributions in this field are in the rheology of zero-Reynolds-number suspensions. Nevertheless, a substantial part of his work during this second period was devoted to other aspects of the subject, including particle dispersion and deposition, mass transfer from particles in linear flows, several aspects of bubble dynamics and fluidized-beds instabilities. This is how he explored and frequently laid the foundations of several branches of modern research in the vast field of two-phase flows. For this, he often relied on the mathematical techniques he developed during the first part of his career devoted to turbulence. His papers, characterized by a unique combination of penetrating physical intuition, mathematical rigor, clarity of exposition and attention to detail remain an inexhaustible source of inspiration. His first contribution to micro-hydrodynamics is now fifty years old. Since then, experimental techniques and computational capabilities have made tremendous progress. However, mathematical models and predictions based on first principles remain the appropriate language to streamline experimental and computational results, and reach a real understanding of the subtle mechanisms at work in complex fluid flows. This is what makes Batchelor’s legacy and conception of research fully alive today. The research presented below seeks to provide a modest illustration of this point of view. \vspace{1mm}\\
\indent Predicting the motion, dispersion and possible accumulation of small rigid particles immersed in nonuniform carrying flows is of paramount importance in all types of two-phase dispersed flows involved in geophysical, biological  and engineering applications. Nowadays, the motion of small spherical particles in nonuniform, possibly turbulent, flows is routinely analyzed through the prism of the Gatignol-Maxey-Riley (GMR) equation \citep{Gatignol1983,Maxey1983}. However, the set of assumptions under which this second-order differential equation for the particle position may be expected to provide a realistic description of the particle fate is quite restrictive. In particular, the particle is assumed to be far from any of its neighbours or from walls, its size has to be small compared to all  characteristic flow length scales, and effects of flow inertia on the particle-induced disturbance have to be negligible, be they due to the particle relative velocity with respect to the carrying flow or to the ambient strain or shear rate.
Consequently, the presence of extra contributions to the hydrodynamic force due to a nearby wall or to the existence of small albeit nonzero flow corrections resulting from fluid inertia are among the effects which are beyond the range of validity of the GMR equation. While the first limitation is presumably clear to everyone, the second is less. Indeed, this equation incorporates some effects of fluid inertia and unsteadiness, namely the so-called added-mass force and the force corresponding to the possible nonzero acceleration of the carrying flow at the position of the particle. However, the contribution of flow inhomogeneity in the Lagrangian fluid acceleration involved in these two forces is generally \textit{not} the leading-order effect due to fluid inertia in the low-but-finite Reynolds regime. This implies that the GMR equation is rarely consistent as soon as fluid inertia comes into play. This is because this contribution to the above two forces is linearly proportional to the particle Reynolds number based on the local shear or strain rate, while leading-order inertial effects in a nonuniform flow are proportional to the square root of this Reynolds number, as exemplified by Saffman's lift force experienced by a small spherical particle translating in a pure shear flow \citep{Saffman1965}.\vspace{2mm}\\
\indent Neutrally-buoyant particles provide an especially stringent test to this equation \citep{Sapsis2011}. Indeed, according to the description it is based upon, the only mechanism capable of producing a velocity difference (so-called slip) between the particle and fluid (assuming that this slip is initially zero) in that case relies on the so-called Faxén force due to the possible curvature of the fluid velocity field at the particle scale. Thus, the GMR equation may for instance correctly predict the longitudinal slip velocity of a neutrally-buoyant particle in a quadratic parallel flow. In contrast, it does not predict any longitudinal slip, nor any lateral migration, when the particle moves in a Couette flow for instance, although it is well-established that both components of slip are nonzero in this case \citep{Halow1970,Ho1974,vasseur1976,Leal1980}. Indeed, small-but-nonzero inertial effects and wall-particle hydrodynamic interactions are at the root of the generation of both slip components in this flow configuration. The same holds true for the transverse migration in a Poiseuille flow.\\
\indent Recently, numerical simulations were performed to explore the dynamics of spherical neutrally-buoyant particles of various sizes released on the axis of an axisymmetric stagnation-point flow, also known as the Hiemenz-Homann flow \citep{Li2019}. This configuration was selected as an archetype of situations in which particles are transported in a flow with a strong wall-normal velocity component, such as that encountered in impinging jets and normal flow filtration, as well as in T-shaped junctions \citep{Vigolo2013}. Numerical results revealed that, starting from zero at large wall-particle distances, the slip velocity becomes increasingly positive as the particle approaches the stagnation point, especially within the boundary layer. This observation indicates that the particle is actually always lagging behind the fluid. However, starting from zero in the bulk (where the flow reduces to a pure bi-axial straining motion), the curvature of the wall-normal velocity component in this flow becomes increasingly negative as the wall is approached. Since the Faxén force is directly proportional to this curvature and the corresponding pre-factor is positive, this force is negative all along the stagnation streamline. Consequently there is no way to explain the generation of a positive slip velocity based on the influence of the Faxén force, hence on the limited physical mechanisms accounted for in the GMR equation (see \S\,\ref{compaDNS} for more discussion). To make the picture unambiguous, it is worth adding that lubrication effects are not the cause of the observed positive slip, as the latter reaches a significant relative magnitude well beyond the separation range within which these effects operate. 
\vspace{2mm}\\
\indent The initial motivation of the present work was provided by the need to rationalize the behaviours revealed by the numerical results of \cite{Li2019}, a goal which could not be reached using the GMR description for the aforementioned reasons. While the inertia-induced migration phenomenon across the flow streamlines has been the subject of many studies over the last half-century in wall-bounded shear flows (see the reviews by \cite{Leal1980} and \cite{Hogg1994}), much less attention has been drawn to wall-normal flows, the archetype of which is the Hiemenz-Homann flow (hereinafter abbreviated as HH flow). The specific configuration in which a sphere is held fixed at a stagnation point was worked out in the creeping-flow limit by \cite{Goren1970}. In the same regime, \cite{Goren1971} considered the case of a sphere moving in the vicinity of a large obstacle held fixed in a streaming flow. This is locally equivalent to the problem of a sphere in motion close to a planar wall with an arbitrary inclination with respect to the upstream flow. Using bi-spherical coordinates, they determined the tangential and wall-normal viscous force and torque components for arbitrary wall-particle gaps, including the range in which lubrication effects are dominant. More recently, \cite{Rallabandi2017} combined the same technique with the use of the reciprocal theorem to develop a comprehensive theory of the viscous forces experienced by a sphere moving along the axis of an axisymmetric wall-normal flow with arbitrary strain and curvature. \\
\indent The aforementioned studies focused on the Stokes-flow regime, disregarding any influence of flow inertia. However these effects can no longer be neglected when the size of the particle increases. In particular, as will be shown later, they become comparable in magnitude with viscous effects when the particle diameter becomes of the order of the boundary layer thickness, which is typical of the situations considered by \cite{Li2019}. To rationalize the trends observed with such `large' neutrally-buoyant particles before their dynamics becomes controlled by lubrication effects, a consistent near-wall force balance incorporating inertial effects is required. The present paper aims at elaborating such a weakly-inertial theory. \\
However, besides helping to rationalize the specific observations of \cite{Li2019}, there is a much broader fundamental interest in providing explicit expressions for near-wall inertial effects in wall-bounded straining flows, which may then be used to predict the particle motion and deposition dynamics in more complex configurations involving a significant wall-normal flow component. To the best of our knowledge, no such theory has been established to date, although the required theoretical tools are available for a long time, especially thanks to the seminal work by \cite{Cox1968}. Considering the three basic kinematic configurations of linear straining, solid-body rotation and uniform shear flows, the latter two are compatible with the presence of a bounding rigid planar wall, provided this wall is parallel to the streamlines of the base flow (\textit{i.e.} perpendicular to the rotation axis in the case of a solid-rotation flow). The situation is more complex in the case of a pure straining motion since such a nonuniform flow cannot satisfy the no-slip condition at the wall. For this reason, a boundary layer within which the vorticity of the base flow is nonzero takes place. It is presumably this more complex structure of the carrying flow that, up to now, hampered the development of a consistent weakly-inertial theory of hydrodynamic forces on a particle in this class of wall-bounded flows. In the spirit of the three fundamental families of linear flows mentioned above, the present work may be seen as the continuation of theoretical investigations such as those of \cite{Cox1977} and \cite{Cherukat1994} for wall-bounded parallel shear flows, or \cite{Magnaudet2003b} (hereinafter referred to as M1) for wall-bounded time-dependent shear and solid-rotation flows. \vspace{2mm}\\
\indent To make the development of such a theory possible, simplifying assumptions are required. 
The reciprocal theorem forms the cornerstone that allows a rigorous force balance to be obtained irrespective of the flow regime. A recent review article \citep{Masoud2019} provides an excellent overview of the amazing variety of low-Reynolds-number transport problems in which this theorem allows the solution to be obtained at a (relatively) low cost. To take advantage of this tool in the present context, we first set up an algebraic approximation of the HH flow yielding an explicit expression of the carrying fluid velocity field down to the wall (\S\,\ref{BLM}). Based on the results derived in M1, the form of the reciprocal theorem suitable to the present problem is re-established in Appendix \ref{appA} and its content is discussed in \S\,\ref{RTH}. Most quantities required to compute explicitly the force contributions revealed by the reciprocal theorem were obtained in M1 and in \cite{Magnaudet2003a} (hereinafter referred to as M2) by solving the so-called `auxiliary' problem with the technique of successive reflections. The corresponding results and their range of validity are summarized in Appendix \ref{CompleA}. Then, guided by the exact force balance offered by the reciprocal theorem, we first derive predictions for the forces acting on a particle released on the flow axis in the creeping-flow limit (\S\,\ref{ZeroRe}).  In a second step, we incorporate inertial corrections, assuming that the Reynolds number is small but finite and the wall-particle separation is small enough for the wall to stand within the inner region of the disturbance (\S\,\ref{inerteff}); details on the procedure used to compute these corrections are provided in Appendix \ref{CIE}. 
Predictions for the particle wall-normal slip velocity  based on the purely viscous force balance and on the improved version incorporating inertial corrections are compared with results from fully-resolved axisymmetric simulations in \S\,\ref{compaDNS} and \S\,\ref{FFB}, respectively. Technical details about these simulations are given in Appendix \ref{Simuli}. Finally we consider the more general configuration where the particle is released at an arbitrary radial position from the stagnation streamline. This configuration, in which the radial and wall-normal particle positions vary over time, represents a fairly general near-wall situation. Indeed, the carrying flow gradually evolves from a pure wall-normal straining motion when the particle stands on the axis of the HH flow, to a pure wall-parallel shear flow when it stands a large distance from the axis. We show that the carrying flow within the boundary layer then comprises a radial shear component producing wall-normal and radial lift contributions, and establish the corresponding force balances on the particle (\S\,\ref{Off}). The main findings of the paper are summarized in \S\,\ref{conclu}.
 \section{Preliminary steps}
 \label{PS}
 \begin{figure}
\vspace{-55 mm}
\hspace{8 mm}{\includegraphics[width=0.99\textwidth]{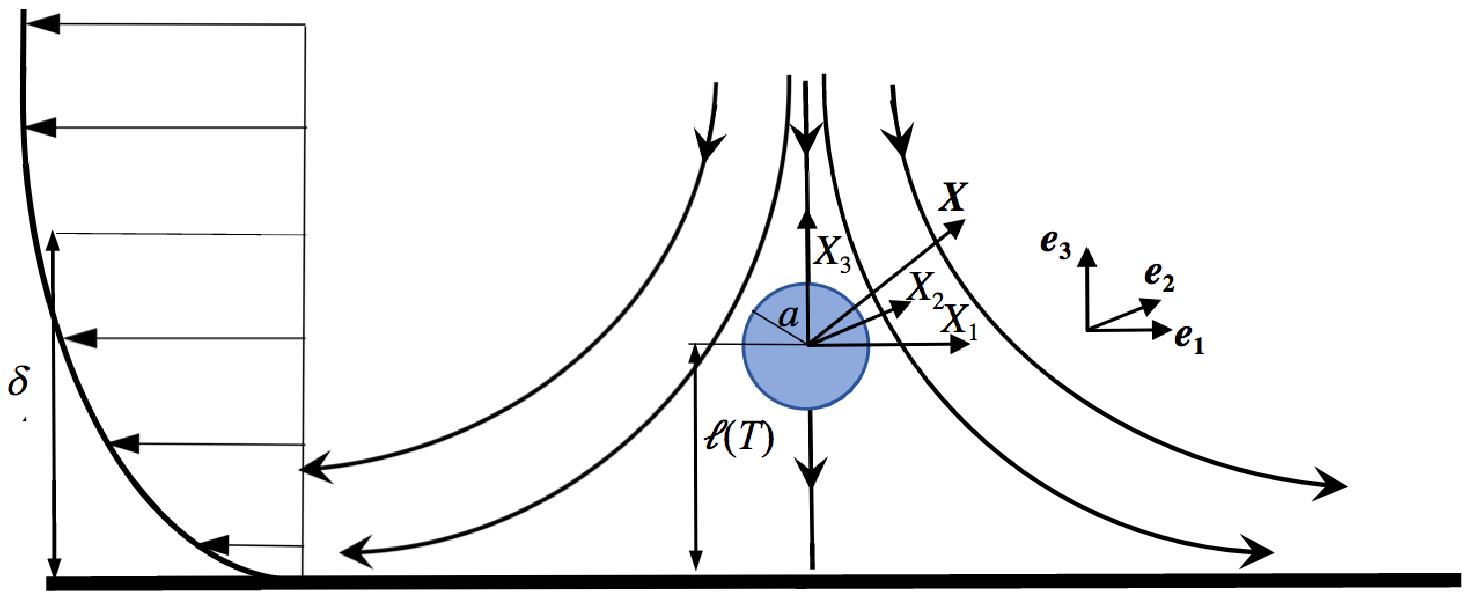}}
\vspace{-88 mm}
\caption[ ]{\footnotesize{Sketch of the flow configuration. The particle radius $a$, wall-particle separation, $\ell$, and boundary layer characteristic thickness, $\delta$, yield the dimensionless length ratios $\kappa=a/\ell$, $\Delta=\delta/a$ and $\Lambda=\delta/k_\delta \ell$ used throughout the paper (the boundary-layer shape parameter $k_\delta$ is defined in \S\,\ref{BLM}).}}
\label{sketch}
\end{figure}
  \subsection{Definitions and scaling}
  \label{scaling}
A Newtonian fluid with uniform density $\rho$ and kinematic viscosity $\nu$ is bounded by a flat wall located in the $(\textbf{\textit{e}}_{\bf{1}},\textbf{\textit{e}}_{\bf{2}})$ plane. The fluid flows towards the wall in the form of an axisymmetric linear straining flow (so-called biaxial straining flow) with a radial (resp. axial) strain rate $B$ (resp. $-2B$). As this inviscid solution does not satisfy the no-slip condition at the wall, a boundary layer with characteristic thickness $\delta=(\nu/B)^{1/2}$ exists along the wall.  We consider a neutrally buoyant spherical particle with radius $a$ standing on the axis of the straining flow and entrained by the fluid towards the wall. At time $T$, the gap between the particle and the wall is $h(T)$, so that the distance separating the particle centre from the wall is $\ell(T)=h(T)+a$ (see figure \ref{sketch}). We make use of a co-ordinate system $\textit{\textbf{X}}=(X_1,X_2,X_3)$ translating with the particle and having its origin at its centre. Then we normalize distances by the particle radius, $a$, whereas time is normalized by a characteristic time scale, $\tau_c$, to be defined later. Velocities are normalized by the unknown slip velocity between the particle and fluid, $V_c$, so that the characteristic Reynolds number is $Re=aV_c/\nu$, the dimensionless  strain rate is $\alpha=aB/V_c$ (hence the product $\alpha Re$ is the strain-based Reynolds number), and forces are normalized by $\rho\nu aV_c$. Beyond the boundary layer, the local fluid velocity with respect to the wall is, in dimensionless form
\begin{equation}
\textbf{\textit{U}}_0(\textbf{\textit{x}},t)\approx\textbf{\textit{U}}_0(\textbf{\textit{x}}={\bf{0}},t)+\alpha({\textbf{\textit{x}}}-3x_3\textbf{\textit{e}}_{\bf{3}})\,,
\label{carflow}
\end{equation}
where $\textbf{\textit{x}}=(x_1,x_2,x_3)=a^{-1}(X_1,X_2,X_3)$ denotes the dimensionless local position with respect to the current position of the particle centre, $t=T/\tau_c$ is the dimensionless time and $\textbf{\textit{e}}_{\bf{3}}$ is the unit normal to the wall directed into the fluid. 
In the momentum balance, the above normalization implies that the advective acceleration is of $\mathcal{O}(Re)$ compared to the viscous term. Similarly, the temporal acceleration is of $\mathcal{O}(ReSt)$, with $St=a/V_c\tau_c$ the Strouhal number comparing the advective time scale $a/V_c$ to the characteristic time $\tau_c$ of the flow. In the specific problem considered here, apart from the possible transient following the release of the particle in the flow, unsteadiness arises because of the non-uniformity of the carrying flow, which transforms into a time-varying flow in the particle reference frame. It is therefore relevant to select $\tau_c=B^{-1}$ as the characteristic time scale, which implies $St\equiv\alpha$. This is why, compared to viscous effects, time-rate-of-change terms are of $\mathcal{O}(\alpha Re)$.\\
\subsection{A rough model for the boundary layer flow}
\label{BLM}
The viscous axisymmetric stagnation point flow problem is governed by a third-order differential equation supplemented by suitable boundary conditions \citep{Homann1936}. Its exact self-similar solution cannot be obtained in closed form and must be determined numerically. To keep the problem tractable analytically, a simple algebraic approximation of this solution is desirable. Rather than trying to fit the full numerical solution with detailed quadratures, we sought a straightforward algebraic divergence-free expression of the velocity field satisfying the no-slip condition at the wall and tending toward (\ref{carflow}) at large distances from it, with a thickness of the transition layer independent from the particle size. Defining the inverse of the dimensionless separation, $\kappa(t)=a/\ell(t)$, we found the simplest base flow satisfying these requirements to be
\begin{equation}
\textbf{\textit{U}}_0(\textbf{\textit{x}},t)=\textbf{\textit{U}}_0(\textbf{\textit{x}}={\bf{0}},t)+\alpha\left\{({\textbf{\textit{x}}}_\parallel-2x_3\textbf{\textit{e}}_{\bf{3}})-\frac{\textbf{\textit{x}}_\parallel}{(1+\mathcal{K}_\delta(\kappa^{-1}+x_3))^2}-\frac{2\mathcal{K}_\delta^{-1}\textbf{\textit{e}}_{\bf{3}}}{1+\mathcal{K}_\delta(\kappa^{-1}+x_3)}\right\}\,,
\label{carflow4}
\vspace{2mm}
\end{equation}
with $\textbf{\textit{x}}_\parallel=x_1\textbf{\textit{e}}_{\bf{1}}+x_2\textbf{\textit{e}}_{\bf{2}}$ and $\mathcal{K}_\delta=k_\delta(\alpha Re)^{1/2}$, $k_\delta$ denoting an adjustable shape parameter to be discussed below. 
\begin{figure}
\centering
{\includegraphics[width=0.6\textwidth]{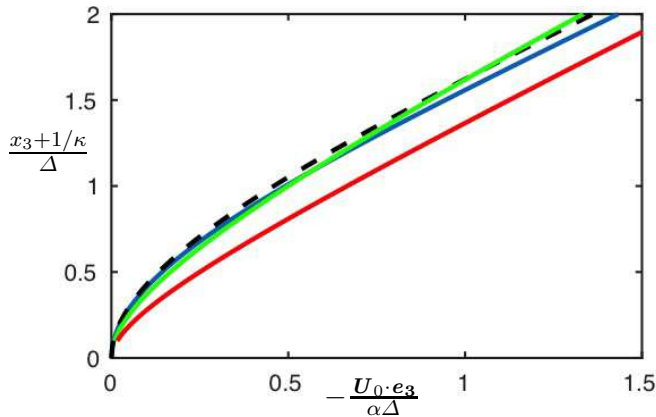}}
\begin{flushleft}
\vspace{-36mm}
\hspace{20mm}\large{$\frac{x_3+1/\kappa}{\Delta}$}\\
\vspace{27mm}\hspace{62mm}\large{$-\frac{\textbf{\textit{U}}_0\cdot\textbf{\textit{e}}_{\bf{3}}}{\alpha\Delta}$}
\end{flushleft}
\vspace{3mm}
\caption[ ]{\footnotesize{Near-wall profile of the wall-normal velocity in the base flow; the velocity and distance to the wall are normalized using boundary layer quantities, \textit{i.e.} $B\delta/V_c=\alpha\Delta$ and $\delta/a=\Delta$, respectively. Blue line: theoretical solution \citep{Homann1936}; dotted line: numerical solution \citep{Li2019}; red and green lines: model (\ref{carflow4}) with $k_\delta=2$ and $k_\delta=1$, respectively.}}
\label{Hiemenz_DNS}
\end{figure}
The first term within braces is the linear straining flow considered in (\ref{carflow}), while the other two contributions represent a rough model of the flow modification within the boundary layer. 
In the reference frame translating with the particle, the wall is located at $x_3=-\kappa^{-1}(t)$. Therefore the no-slip condition $\textbf{\textit{U}}_0(\textbf{\textit{x}}_\parallel,x_3=-\kappa^{-1},t)={\bf{0}}$ implies that the fluid velocity at the current position of the particle centre is $\textbf{\textit{U}}_0(\textbf{\textit{x}}={\bf{0}},t)=2\alpha(\mathcal{K}_\delta^{-1}-\kappa^{-1})\textbf{\textit{e}}_{\bf{3}}$. \\
Since $\alpha Re=a^2B/\nu\equiv a^2/\delta^2$, the dimensionless characteristic boundary layer thickness $\Delta$ obeys the relation $\Delta=(\alpha Re)^{-1/2}$, which implies $\mathcal{K}_\delta=k_\delta\Delta^{-1}$. Hence the second term within curly braces in (\ref{carflow4}) reduces to $-\textbf{\textit{x}}_\parallel(1+k_\delta)^{-2}$ when the particle stands a distance $\kappa^{-1}=\Delta$ from the wall. 
With $k_\delta=2$, the tangential velocity $\alpha \textbf{\textit{x}}_\parallel\left(1-(1+k_\delta)^{-2}\right)$ reaches approximately $90\%$ of its free-stream value at this position, a percentage that increases to $98\%$ for $\kappa^{-1}=3\Delta$. These features are in good agreement with the actual velocity profile of the HH flow displayed in figure 2 of \cite{Li2019}. Thus (\ref{carflow4}) with $k_\delta\approx2$ is expected to represent well the variation of the carrying flow in the part of the boundary layer close to its outer edge. 
However, the approximate base flow must also correctly estimate the curvature $\mathcal{C}$ of the normal velocity $\textbf{\textit{U}}_0\cdot\textbf{\textit{e}}_{\bf{3}}$ in the limit $x_3\rightarrow-1/\kappa$, since this curvature governs the variation of all three velocity components within the inner part of the boundary layer, say for $0\le x_3+1/\kappa\lesssim\Delta$. In this limit, the velocity field (\ref{carflow4}) reduces to the nearly-parallel distribution $\textbf{\textit{U}}_0(\textbf{\textit{x}},t)\approx2\mathcal{K}_\delta\alpha(\kappa^{-1}+x_3)\{{\textbf{\textit{x}}}_\parallel-(\kappa^{-1}+x_3)\textbf{\textit{e}}_{\bf{3}}\}$, so that (\ref{carflow4}) predicts $\mathcal{C}\approx-4\mathcal{K}_\delta\alpha=-4k_\delta\alpha/\Delta$. Figure \ref{Hiemenz_DNS} shows how this model approaches the variation of $\textbf{\textit{U}}_0\cdot\textbf{\textit{e}}_{\bf{3}}$ encountered near the wall in the actual HH flow. It turns out that the above value $k_\delta=2$ significantly overestimates $\mathcal{C}$, hence $-\textbf{\textit{U}}_0\cdot\textbf{\textit{e}}_{\bf{3}}$, throughout this region and even beyond. A much better agreement with the actual profile is obtained with $k_\delta=1$. Nevertheless, with this lower $k_\delta$, the tangential velocity reaches $98\%$ of its free-stream value only for $\kappa^{-1}=6\Delta$. Hence it appears that a single value of $k_\delta$ does not allow (\ref{carflow4}) to fit closely the actual near-wall flow throughout the boundary layer. This is not unexpected since the velocity field in (\ref{carflow4}) is not an exact solution of the Navier-Stokes equation. Indeed, the corresponding vorticity, $\boldsymbol\omega_\delta(\textbf{\textit{x}})=-2\alpha\mathcal{K}_\delta(x_2\textbf{\textit{e}}_{\bf{1}}-x_1\textbf{\textit{e}}_{\bf{2}})(1+\mathcal{K}_\delta(\kappa^{-1}+x_3))^{-3}$, does not satisfy the vorticity transport equation, except in the region closest to the wall ($\kappa^{-1}+x_3\ll1$). 
Nevertheless, since the influence of boundary layer effects on the particle dynamics is expected to be large essentially within the $\mathcal{O}(\Delta)$-thick region next to the wall, it is likely that $k_\delta=1$ is the optimal choice to be used in conjunction with the simple model (\ref{carflow4}). Comparisons of slip velocities predicted by the present theory with results of fully-resolved simulations will later confirm this conclusion (see figure \ref{Slip_1}$(b)$). However, to keep the results more general, $k_\delta$ will be left unspecified throughout the developments performed in the next sections.\vspace{2mm}\\
 \indent Returning to (\ref{carflow4}) and defining 
 \begin{equation}
 \textbf{\textit{U}}_{0}^0(t)=\textbf{\textit{U}}_0(\textbf{\textit{x}}={\bf{0}},t)\quad \mbox{and}\quad\Lambda(t)=\frac{\kappa(t)}{\mathcal{K}_\delta}=\frac{\kappa(t)\Delta}{k_\delta}\,, 
 \label{Defin}
 \end{equation}
 the carrying flow close to the particle (formally within the region $|x_3| \ll(1+\Lambda)/\kappa$) may be expanded in the form
\begin{equation}
\textbf{\textit{U}}_0(\textbf{\textit{x}},t)=\textbf{\textit{U}}_0^0(t)+\alpha_b(t)({\textbf{\textit{x}}}-3x_3\textbf{\textit{e}}_{\bf{3}})+\alpha_c(t)x_3({\textbf{\textit{x}}}-2x_3\textbf{\textit{e}}_{\bf{3}})+...\,,
\label{carflow5}
\end{equation}
with 
\begin{equation}
\textbf{\textit{U}}_0^0(t)=-2\alpha\frac{1}{\kappa(t)(1+\Lambda(t))}\textbf{\textit{e}}_{\bf{3}}\,,\quad \alpha_b(t)=\alpha\frac{1+2\Lambda(t)}{(1+\Lambda(t))^2}\,,\quad \alpha_c(t)=2\alpha\kappa(t)\frac{\Lambda^2(t)}{(1+\Lambda(t))^3}\,.
\label{coeff}
\vspace{1.5mm}
\end{equation}
The inviscid base flow (\ref{carflow}) is recovered in the limit $\Lambda\rightarrow0$, for which $\alpha_b\rightarrow\alpha$ and $\alpha_c\rightarrow0$. 
For finite $\Lambda$, the leading influence of the boundary layer is to reduce the effective strain rate at the position of the particle to an $\mathcal{O}((1+2\Lambda)/(1+\Lambda)^2)$-fraction of its free-stream value, and to introduce a quadratic component of the flow with an $\mathcal{O}(\kappa\Lambda^2/(1+\Lambda)^3)$-magnitude. The quantity $\Lambda^{-1}=k_\delta(\kappa\Delta)^{-1}$ may be thought of as the distance separating the particle from the wall normalized by the effective boundary layer thickness $6\Delta/k_\delta$, the distance to the wall at which the tangential velocity reaches $98\%$ of its free-stream value. For reasons to be discussed later, the asymptotic approach developed in the next sections will be restricted to particles much smaller than the boundary layer thickness, which implies $\Delta\gg1$.  For such particles, $\Lambda$ varies from near-zero values when the particle is far from the boundary layer $(\kappa\rightarrow0$) to large $\mathcal{O}(\Delta)$-values (since $1\lesssim k_\delta\lesssim2$) when it gets very close to the wall. 

 \subsection{Reciprocal theorem}
 \label{RTH}
Forces acting on a spherical buoyant drop with an arbitrary viscosity immersed in a linear flow bounded by a single flat wall and translating with velocity $\textbf{\textit{V}}$ in an arbitrary direction with respect to that wall were considered in M1. In a preliminary step, a general expression for the force balance, valid whatever the magnitude of unsteadiness and inertia effects, was obtained by making use of the reciprocal theorem. It is straightforward to extend this force balance to the quadratic flow (\ref{carflow5}), and consider the particular case of a neutrally-buoyant rigid particle. For the sake of self-consistency, the main steps of the derivation are provided in Appendix \ref{appA}. As is well known, evaluating wall-normal forces with the help of the reciprocal theorem requires the determination of the solution of the `auxiliary' problem corresponding to a spherical particle translating perpendicularly to the wall with unit velocity in a fluid at rest. Let $\hat{\textbf{\textit{U}}}$ and $\hat{\boldsymbol{\Sigma}}$ be the fluid velocity and stress fields associated with this problem, respectively. Then let $\textbf{\textit{u}}(\textbf{\textit{x}},t)$ and $\textbf{\textit{V}}_{S0}(t)=\textbf{\textit{V}}(t)-\textbf{\textit{U}}_0^0(t)$ be the velocity disturbance and time-dependent slip velocity between the particle and fluid involved in the actual (`direct') problem, respectively. 
Using the scalings established in \S\,\ref{scaling}, the derivation in Appendix A provides the exact dimensionless force balance on a rigid neutrally buoyant spherical particle moving perpendicular to the wall in the form (\ref{recipr11}). This result being valid for an arbitrary carrying flow, the force balance in a quadratic flow such as that defined by (\ref{carflow5}) becomes
\begin{eqnarray}
\nonumber
\nonumber
&&Re\left(\frac{4}{3}\pi\alpha \frac{d\textbf{\textit{V}}}{dt}-\int_\mathcal{V_A}\frac{D\textbf{\textit{U}}_0}{Dt}d\mathcal{V}\right)\cdot\textbf{\textit{e}}_{\bf{3}}=\hat{\textbf{\textit{F}}}_D\cdot\textbf{\textit{V}}_{S0}-\hat{\textbf{\textit{T}}}_D:\nabla^0\textbf{\textit{U}}_0-\frac{1}{2}\hat{\textbf{\textit{S}}}_D\scalebox{.7 }[0.7 ]{\vdots}\nabla^0\nabla{\textbf{\textit{U}}_0}\\
&&
-Re\int_\mathcal{V}(\hat{\textbf{\textit{U}}}+\textbf{\textit{e}}_{\bf{3}})\cdot\left(\alpha\frac{\partial\textbf{\textit{u}}}{\partial t}+\textbf{\textit{u}}\cdot\nabla\textbf{\textit{U}}_0+(\textbf{\textit{U}}_0-\textbf{\textit{U}}_0^0)\cdot\nabla\textbf{\textit{u}}+(\textbf{\textit{u}}-\textbf{\textit{V}}_{S0})\cdot\nabla\textbf{\textit{u}}\right)d\mathcal{V}\,,\quad\quad
\label{recipr}
\end{eqnarray}
where $\mathcal{V_A}$ and $\mathcal{V}$ refer to the volume occupied by the particle and the fluid, respectively, and $\hat{\textbf{\textit{F}}}_D=\int_\mathcal{A}\hat{\boldsymbol{\Sigma}}\cdot\textbf{\textit{n}}d\mathcal{A}$ is the drag force on the particle in the auxiliary problem, $\textbf{\textit{n}}$ denoting the unit normal to the particle surface $\mathcal{A}$ directed into the fluid. The gradient $\nabla^0\textbf{\textit{U}}_0=\nabla\textbf{\textit{U}}_0(\textbf{\textit{x}}=\textbf{0})=\alpha_b(\textbf{\textit{I}}-3\textbf{\textit{e}}_{\bf{3}}\textbf{\textit{e}}_{\bf{3}})$ and Hessian $\nabla^0\nabla\textbf{\textit{U}}_0=\nabla(\nabla\textbf{\textit{U}}_0)(\textbf{\textit{x}}=\textbf{0})=\alpha_c\textbf{\textit{e}}_{\bf{3}}(\textbf{\textit{I}}-2\textbf{\textit{e}}_{\bf{3}}\textbf{\textit{e}}_{\bf{3}})$ of the undisturbed velocity (\ref{carflow5}) at the centre of the particle being nonzero, they provide additional contributions to the force through the first- and second-order moments of the surface stress, $\hat{\textbf{\textit{T}}}_D=\int_\mathcal{A}\textbf{\textit{x}}\hat{\boldsymbol{\Sigma}}\cdot\textbf{\textit{n}}d\mathcal{A}$ and $\hat{\textbf{\textit{S}}}_D=\int_\mathcal{A}\textbf{\textit{x}}\textbf{\textit{x}}\hat{\boldsymbol{\Sigma}}\cdot\textbf{\textit{n}}d\mathcal{A}$, with $\textbf{\textit{x}}$ the local position with respect to the particle centre and $\textbf{\textit{I}}$ the Kronecker delta. In (\ref{recipr}), $d/dt$ is the time derivative following the particle motion, while $D\textbf{\textit{U}}_0/Dt$ is the acceleration of the undisturbed carrying flow. In the reference frame translating with the particle, this acceleration reads $D\textbf{\textit{U}}_0/Dt=\alpha d\textbf{\textit{U}}_0/dt+(\textbf{\textit{U}}_0-\textbf{\textit{V}})\cdot\nabla\textbf{\textit{U}}_0$, the $\alpha$-pre-factor resulting from the scaling of unsteady effects as discussed in \S\,\ref{scaling}. \\
Beyond the boundary layer, the carrying flow is linear, implying $\nabla^0\nabla{\textbf{\textit{U}}}_0=\textbf{0}$ and making the undisturbed fluid acceleration uniform, hence the left-hand side of (\ref{recipr}) proportional to the relative acceleration $\alpha d\textbf{\textit{V}}/dt-D\textbf{\textit{U}}_0/Dt$. Since $\alpha_b=\alpha$, $\nabla\textbf{\textit{U}}_0=\alpha(\textbf{\textit{I}}-3\textbf{\textit{e}}_{\bf{3}}\textbf{\textit{e}}_{\bf{3}})$ is of ${\mathcal{O}}(\alpha)$ there, and all terms in (\ref{recipr}) involving the fluid and particle accelerations are of $\mathcal{O}(\alpha Re)$. The left-hand side of (\ref{recipr}) then yields a net inertial force $\textbf{\textit{F}}_0$ on the particle 
\begin{equation}
\textbf{\textit{F}}_0\cdot{\textbf{\textit{e}}_{\bf{3}}}=\frac{4}{3}\pi \alpha Re\left(\frac{d\textbf{\textit{V}}_{S0}}{dt}-2\textbf{\textit{V}}_{S0}\right)\cdot{\textbf{\textit{e}}_{\bf{3}}}\,.
\label{F0}
\end{equation}
Within the boundary layer, the local strain rates $\alpha_b(t)$ and $\alpha_c(t)$ in (\ref{carflow5}) vary with the position of the particle with respect to the wall. Then an additional force proportional to $\alpha_c(t)$ takes place, owing to the $-\frac{1}{2}\hat{\textbf{\textit{S}}}_D\scalebox{.7 }[0.7 ]{\vdots}\nabla^0\nabla{\textbf{\textit{U}}_0}$ contribution. Moreover, the body force $\int_\mathcal{V_A}\frac{D\textbf{\textit{U}}_0}{Dt}d\mathcal{V}$ includes quadratic corrections proportional to $d\alpha_c/dt$ and $\alpha_b(t)\alpha_c(t)$ which modify (\ref{F0}) into
\begin{equation}
\textbf{\textit{F}}_0\cdot{\textbf{\textit{e}}_{\bf{3}}}=\frac{4}{3}\pi Re \left\{\left(\alpha\frac{d\textbf{\textit{V}}_{S0}}{dt}-2\alpha_b\textbf{\textit{V}}_{S0}\right)\cdot{\textbf{\textit{e}}_{\bf{3}}}+\frac{1}{5}\left(6\alpha_b\alpha_c-\frac{d\alpha_c}{dt}\right)\right\}\,.
\label{F0BL}
\end{equation}
\subsection{Solving the auxiliary problem}
\label{Comple0}
To make practical use of (\ref{recipr}), a key step is to solve the auxiliary problem. An exact solution of this problem based on bipolar co-ordinates, valid until the particle touches the wall, was derived independently by \cite{Brenner1961} and \cite{Maude1961}. Nevertheless making use of the corresponding solution to compute inertial terms involved in the right-hand side of (\ref{recipr}) is nontrivial. A more tractable approach consists in assuming formally that the separation between the particle and the wall is large and seeking the solution in the form of a series of `reflections' of the fundamental solution corresponding to a particle translating in an unbounded fluid. To this end, it is customary to expand the solution with respect to the small parameter $\kappa=a/\ell=(1+\epsilon)^{-1}$, where $\epsilon=h/a$ is the dimensionless gap. An approximate solution truncated at $\mathcal{O}(\kappa^4)$ was obtained in M1 and M2 using this technique. The main steps involved in the elaboration of this solution are summarized in Appendix \ref{CompleA}, together with the explicit expressions for $\hat{\textbf{\textit{F}}}_D$, $\hat{\textbf{\textit{T}}}_D$ and $\hat{\textbf{\textit{S}}}_D$ required to evaluate the first three contributions in the right-hand side of (\ref{recipr}). This appendix also discusses the limit of validity of this approximate solution, determined by comparing its predictions for the drag force with exact solutions and computational results. The conclusion is that this truncated solution is valid approximately up to $\kappa=0.5$, \textit{i.e.} down to $\epsilon\approx1$. Clearly, lubrication effects that take place when $\kappa\rightarrow1$ ($\epsilon\rightarrow0$) cannot be captured and stay beyond the capabilities of the present asymptotic theory.
\section{Zero-Reynolds-number approximation}
 \label{ZeroRe}

We now assume that inertia effects are small, \textit{i.e.} $Re\ll1$ and $\alpha Re\ll1$. Actually, since the particle is considered to be neutrally buoyant, the dimensional slip velocity $V_c$ is expected to be much smaller than the strain-based velocity $Ba$, so that $\alpha$ is large. Hence the previous two conditions may be ordered in the form
\begin{equation}
Re\ll \alpha Re\ll 1\,.
\label{cond1}
\end{equation}
However, $\alpha Re=a^2B/\nu$ and $B\delta^2/\nu=1$ by definition, so that the dimensionless characteristic boundary layer thickness $\Delta=\delta/a$ is such that $\Delta=(\alpha Re)^{-1/2}$. Hence (\ref{cond1}) may be rewritten in the form 
\begin{equation}
Re\ll \Delta^{-2}\ll 1\,.
\label{cond3}
\end{equation}
This condition implies that for the strain Reynolds number $\alpha Re$ to be small, the particle must be much smaller than the boundary layer thickness. This is why only `small' particles satisfying this condition fall into the field of application of the asymptotic theory developed in the rest of this paper.
\subsection{Wall- and curvature-induced Faxén forces}
\indent In this section we totally disregard inertial effects, which in particular implies that the contributions of the volume integrals in the left- and right-hand sides of (\ref{recipr}) are neglected. The total force acting on the particle is then merely the sum of the contributions resulting from the slip velocity $\textbf{\textit{V}}_{S0}$, and the successive gradients of the carrying flow at the position of the particle, $\nabla^0{\textbf{\textit{U}}_0}$ and $\nabla^0\nabla{\textbf{\textit{U}}_0}$.\vspace{2mm} \\
\indent Inserting the explicit expression for $\hat{\textbf{\textit{T}}}_D$ provided by (\ref{comple2}) in (\ref{recipr}), with $\nabla^0{\textbf{\textit{U}}_0}$ derived from (\ref{carflow5}), reveals that in the present axisymmetric straining flow the force moment $\hat{\textbf{\textit{T}}}_D=\int_\mathcal{A}\textbf{\textit{x}}(\hat{\boldsymbol{\Sigma}}\cdot{\textbf{\textit{e}}_r})d\mathcal{S}$ yields a net force on the particle
\begin{equation}
{\textbf{\textit{F}}}_{F}\cdot\textbf{\textit{e}}_{\bf{3}}=\frac{45}{4}\pi\alpha_b\kappa^2(1+\frac{9}{8}\kappa+...)\,.
\label{faxen}
\end{equation}
This force tends to repel the particle from the wall, \textit{i.e.} to make it lag behind the impinging straining flow (\ref{carflow}). With reference to the well-known Faxén force resulting from the inhomogeneity of the undisturbed velocity field in quadratic flows,  this contribution may be thought of as a wall-induced Faxén force. Its origin is made clear by considering the fundamental solution of the `direct' problem in the unbounded case. As the particle is neutrally buoyant, this solution is merely the sum of a stresslet and an irrotational  quadrupole. Since the disturbance induced by the stresslet decays as $r^{-2}$, with $r=||\textbf{\textit{x}}||$ the distance to the particle centre, its reflection on the wall induces a velocity correction proportional to $\alpha_b\kappa^2\bf{e}_3$ in the vicinity of the particle, yielding an $\mathcal{O}(\kappa^2)$-repelling force. \cite{Rallabandi2017} made use of bipolar co-ordinates to evaluate the drag force acting on a spherical particle translating perpendicularly to a curved wall along the axis of an arbitrary nonuniform axisymmetric flow. They found that the linear variation of the flow induces a normal force, say $\textbf{\textit{F}}_{RA}\cdot\textbf{\textit{e}}_{\bf{3}}$, which in present notations reads $-6\pi\mathcal{B}\textbf{\textit{e}}_{\bf{3}}\cdot\nabla^0\textbf{\textit{U}}_0\cdot\textbf{\textit{e}}_{\bf{3}}$. In the limit of large gaps and weak wall curvature, $\mathcal{B}\rightarrow\frac{15}{16}\epsilon^{-2}$ (their equation (5.4$a$)). Since $\kappa\approx\epsilon^{-1}$ in that limit and $\textbf{\textit{e}}_{\bf{3}}\cdot\nabla^0\textbf{\textit{U}}_0=-2\alpha_b\textbf{\textit{e}}_{\bf{3}}$ in the present flow, their result may be re-written in the form $\textbf{\textit{F}}_{RA}\cdot\textbf{\textit{e}}_{\bf{3}}\rightarrow\frac{45}{4}\pi\kappa^{2}\alpha_b$ in this specific situation, which is exactly the leading-order contribution in (\ref{faxen}). For $\epsilon=1$ ($\kappa=1/2$), the $\mathcal{O}(\kappa^3)$-approximation of ${\textbf{\textit{F}}}_{F}$ provided by (\ref{faxen}) and the exact solution of \cite{Rallabandi2017} differ by less than $13\%$. \vspace{2mm}\\
Evaluating now the contribution of the quadratic flow component $\nabla^0\nabla{\textbf{\textit{U}}_0}$ in (\ref{recipr}) with the aid of (\ref{comple3}), we find that the corresponding force is 
 \begin{equation}
  {\textbf{\textit{F}}_{F\delta}}\cdot\textbf{\textit{e}}_{\bf{3}}=\pi(1+\frac{9}{8}\kappa+\frac{81}{64}\kappa^2+\frac{217}{512}\kappa^3)(\nabla^2)^0{\textbf{\textit{U}}_0}\cdot\textbf{\textit{e}}_{\bf{3}}+\frac{15}{8}\pi\kappa^3\textbf{\textit{e}}_{\bf{3}}\cdot\nabla^0(\textbf{\textit{e}}_{\bf{3}}\cdot\nabla{\textbf{\textit{U}}_0})\cdot\textbf{\textit{e}}_{\bf{3}}+\mathcal{O}(\kappa^4)\,,
 \label{Facla}
 \end{equation}
where $(\nabla^2)^0{\textbf{\textit{U}}_0}$ denotes the Laplacian of the carrying velocity field at the position of the particle centre. The corresponding term in (\ref{Facla}) is the classical Faxén force originating in the curvature of the carrying flow. In the present context, this force is zero when the particle stands in the outer flow region, but increases as it approaches the wall once it is immersed within the boundary layer. A similar force component was computed by \cite{Rallabandi2017} who, in present notations, wrote it in the form $3\pi\mathcal{D}(\nabla^2)^0{\textbf{\textit{U}}_0}\cdot\textbf{\textit{e}}_{\bf{3}}$. Figure 3 in their paper indicates that $\mathcal{D}\rightarrow1/3$ for $\kappa\rightarrow0$ and increases to $0.65$ for $\kappa=1/2$. The prediction (\ref{Facla}) fully agrees with this variation, with less than $1\%$ difference for $\kappa=1/2$. 
The contribution proportional to $\nabla^0(\textbf{\textit{e}}_{\bf{3}}\cdot\nabla{\textbf{\textit{U}}_0})\cdot\textbf{\textit{e}}_{\bf{3}}$ in (\ref{Facla}) results from the anisotropy introduced by the wall at $\mathcal{O}(\kappa^3)$ in the solution of the auxiliary problem (see the discussion in Appendix \ref{CompleA}). The force resulting from this contribution was also computed by \cite{Rallabandi2017} ($\mathcal{C}$-term in their equation (4.12) and figure 3). In the present context, the quadratic velocity component in (\ref{carflow5}) is of $\mathcal{O}(\alpha_c)$, hence of $\mathcal{O}(\kappa)$ for a given $\Lambda$ according to (\ref{coeff}), so that the $\mathcal{O}(\kappa^3)$-terms in (\ref{Facla}) have to be neglected to remain consistent with the general $\mathcal{O}(\kappa^4)$-truncation discussed in \S\,\ref{Comple0}. 
   With ${\textbf{\textit{U}}_0}$ given by (\ref{carflow5}), (\ref{Facla}) then yields 
 \begin{equation}
 {\textbf{\textit{F}}_{F\delta}}\cdot\textbf{\textit{e}}_{\bf{3}}\approx-2\pi(1+\frac{9}{8}\kappa+\frac{81}{64}\kappa^2)\alpha_c\,.
 \label{Faclaa}
 \end{equation}
 \vspace{1mm}\\
 \noindent Finally, taking into account (\ref{comple1}), (\ref{faxen}) and (\ref{Faclaa}) and the definitions of $\alpha_b$ and $\alpha_c$ in (\ref{coeff}), the zero-$Re$ force balance resulting from (\ref{recipr}) is found to be
 \begin{equation}
 24(1+\frac{9}{8}\kappa+...)\textbf{\textit{V}}_{S0}\cdot\textbf{\textit{e}}_{\bf{3}}\approx\alpha\kappa\left\{45\frac{1+2\Lambda}{(1+\Lambda)^2}(1+\frac{9}{8}\kappa+...)\kappa-16(1+\frac{9}{8}\kappa+...)\frac{\Lambda^2}{(1+\Lambda)^3}\right\}\,.
 \label{Rezerod}
 \end{equation}
The wall-induced force (\ref{faxen}) resulting from the gradients of the carrying flow is responsible for the first contribution within the curly brackets. It tends to produce a positive slip velocity growing quadratically as the separation decreases. The curvature-induced Faxén force (second term within the curly brackets) acts to reduce this positive slip. However the resulting behaviour is not entirely intuitive. In the limit of large separations, \textit{i.e.} $\Lambda\rightarrow0$, the right-hand side of (\ref{Rezerod}) is positive only if $\kappa\lesssim\frac{45}{16}k_\delta^2\Delta^{-2}(1-\frac{135}{16}k_\delta\Delta^{-1})^{-1}$. So, at a given separation such that $\kappa\ll\Delta^{-1}$, only sufficiently large particles experience a positive slip. 
 For instance, with $k_\delta=1$, the slip of a particle 20 times smaller than the boundary layer characteristic thickness (\textit{i.e.} such that $\Delta=20$) is found to be positive for $\kappa\lesssim0.014$ but is then negative until $\kappa\approx0.089$ before it becomes positive again for smaller separations. 
 Very close to the wall, $\Lambda$ is large for small particles. Therefore both terms in the right-hand side of (\ref{Rezerod}) behave as $1/\Lambda$ in that limit but the large pre-factor of the first of them ensures that the positive driving force dominates. For instance, still with $k_\delta=1$, $\Lambda=2.5$ (resp. $5$) when $\kappa=1/2$ (resp. $1$) for particles corresponding to $\Delta=5$, so that the positive force is approximately $4.5$ (resp. $7.5$) times larger than the negative one. That the slip velocity predicted by (\ref{Rezerod}) is positive whatever the particle size in the limit $\kappa\rightarrow1$ is of physical interest, although the present theory is not expected to apply in that limit. Since the fluid velocity is still negative (\textit{i.e.} directed towards the wall) at the position of the particle centre, but the velocity of the particle has to vanish when the latter touches the wall, the actual slip velocity is undoubtedly positive. Obviously, lubrication effects not accounted for in the present theory contribute to slow down the particle as it gets very close to the wall \citep{Li2019}. Nevertheless, what (\ref{Rezerod}) reveals is that the longer-range hydrodynamic forces considered here contribute to this slowing down, as they force the slip velocity to be positive and to increase with $\kappa$ for $\kappa\lesssim1$.
 \subsection{Comparison with numerical results}
\label{compaDNS}
\begin{figure}
\centering
\vspace{2mm}
\begin{subfigure}[]
\centering
{\includegraphics[width=0.486\textwidth]{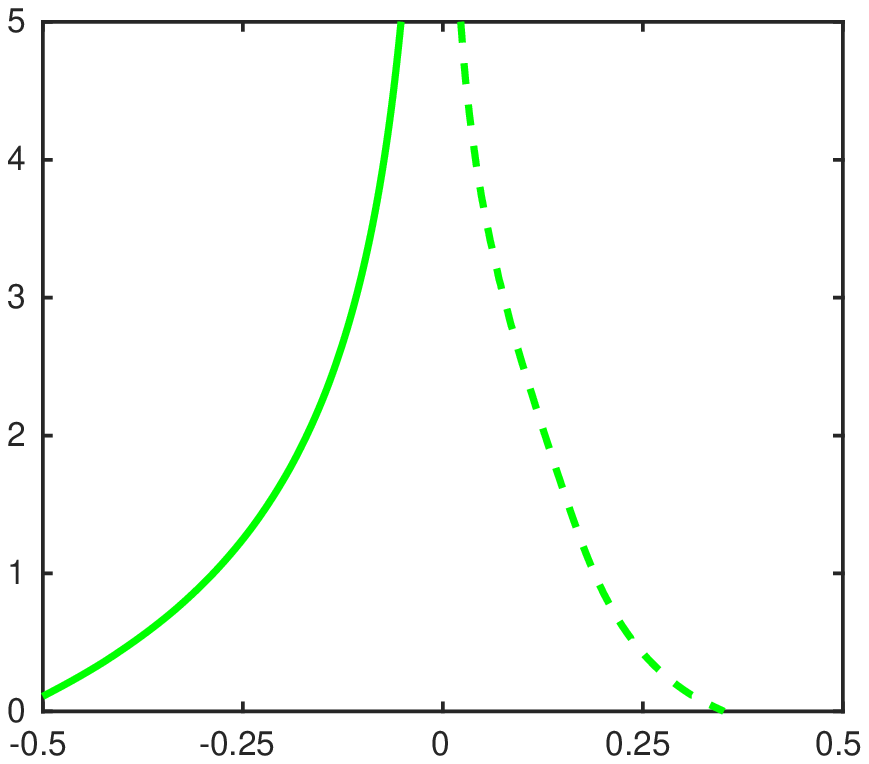}}
    \end{subfigure} 
    \hfill
\begin{subfigure}[]
\centering
{\includegraphics[width=0.494\textwidth]{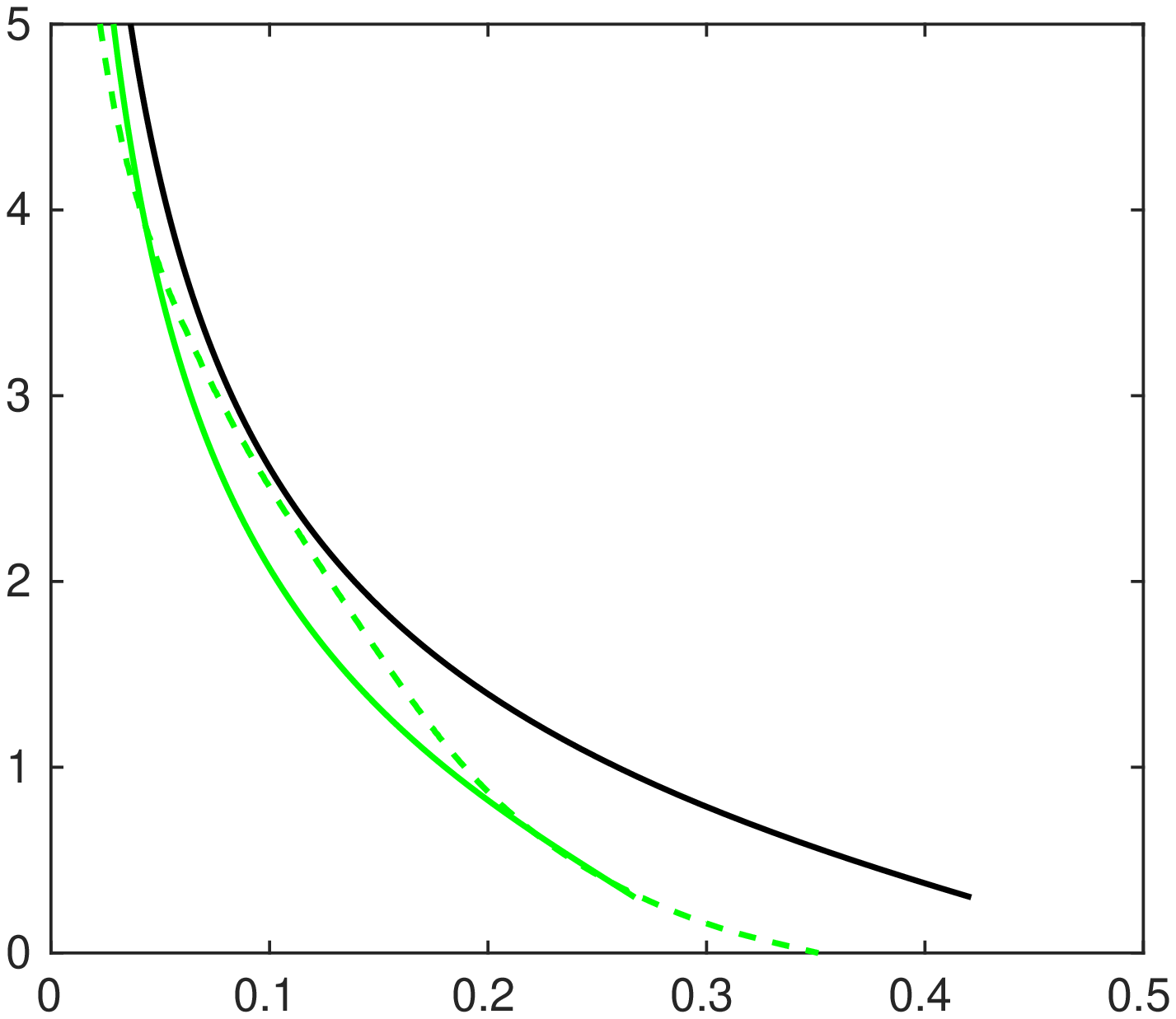}}
   \end{subfigure}
       \hfill
       \begin{flushleft}
\vspace{-46.5mm}\hspace{4mm}$\epsilon$\hspace{67mm}$\epsilon$\\
\vspace{33.8mm}
\hspace{25mm}$\alpha^{-1}\textbf{\textit{V}}_{S0}\cdot\textbf{\textit{e}}_{\bf{3}}$\hspace{50.5mm}$\alpha^{-1}\textbf{\textit{V}}_{S0}\cdot\textbf{\textit{e}}_{\bf{3}}$
\end{flushleft}
\vspace{2mm}
\caption{Slip velocity profile as a function of the gap $\epsilon=\kappa^{-1}-1$ for a particle with relative radius $\Delta^{-1}=0.3$ compared to the characteristic boundary layer thickness. $(a)$ Comparison between simulation results (\protect\greendashline) and predictions of the GMR equation using the undisturbed flow (\ref{carflow4}) with $k_\delta=1$ (\protect\greenline); $(b)$ comparison between simulation results (\protect\greendashline) and predictions of (\ref{Rezerod}) with $k_\delta=2$ (\protect\blackthinline) and $k_\delta=1$ (\protect\greenline).}
\label{Slip_1}
\end{figure}
\begin{figure}
\centering
\vspace{2mm}
\includegraphics[width=0.55\textwidth]{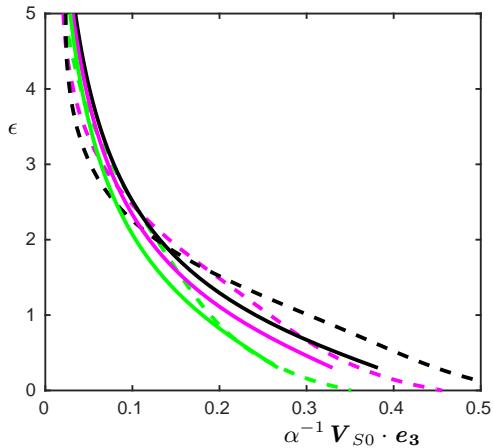}
\begin{flushleft}
\vspace{-44mm}\hspace{35mm}$\epsilon$\\
\vspace{37mm}
\hspace{71.5mm}$\alpha^{-1}\textbf{\textit{V}}_{S0}\cdot\textbf{\textit{e}}_{\bf{3}}$
\end{flushleft}
\vspace{0mm}
\caption{Slip velocity in the near-wall region for particles with increasing relative size $\Delta^{-1}=0.3$ (\protect\greenline), $0.4$ (\protect\magentaline), $0.5$ (\protect\blackline). 
Dashed line: simulation results; thin solid line: creeping-flow prediction (\ref{Rezerod}) using the undisturbed flow (\ref{carflow4}) with $k_\delta=1$.}
\label{Slip_prof0}
\end{figure}

\cite{Li2019} reported results of fully-resolved numerical simulations carried out with particles released from rest on the stagnation streamline of a HH flow. 
Although analyses in their paper focus on `large' particles, some of which with radii of the order of the total boundary layer total thickness (up to $\Delta^{-1}=3.2$), other simulations were run with smaller particles, corresponding to relative sizes $\Delta^{-1}$ down to $0.1$ (Li 2019, private communication). Technical details about these simulations are provided in Appendix \ref{Simuli}. Here we select some of these results obtained with `small' particles to discuss several features of the near-wall variations of the slip velocity $\textbf{\textit{V}}_{S0}$ with the position of the particle, and compare present zero-Reynolds-number predictions (which are in principle only valid for $\Delta^{-1}\ll1$) with those of the full Navier-Stokes equations. 
 In figures \ref{Slip_1}-\ref{Slip_prof}, slip profiles are plotted \textit{vs.} the dimensionless gap $\epsilon=\kappa^{-1}-1$ to make  the physical interpretation easier.\\
 \indent First of all, figure \ref{Slip_1}$(a)$ compares the numerical slip velocity profile typical of a small particle (with a radius ten times smaller that the total boundary layer thickness $3\Delta$) with the prediction of the GMR model. In this case, the strain Reynolds number is $0.09$ and the maximum slip-based Reynolds number is less than $0.03$, so that inertial effects are expected to be negligibly small throughout the particle trajectory. Hence the GMR model (\textit{e.g.} equation (48) in \cite{Maxey1983}) reduces to a balance between the viscous drag linearly proportional to $\textbf{\textit{V}}_{S0}$ and the curvature-induced Faxén force proportional to $(\nabla^2)^0\textbf{\textit{U}}_0$, both of which evaluated as if the particle motion were taking place in an unbounded fluid. In the notations of (\ref{recipr}), this balance results in 
 \begin{equation}
 \hat{\textbf{\textit{F}}}_D^\infty\cdot\textbf{\textit{V}}_{S0}\approx\frac{1}{2}\hat{\textbf{\textit{S}}}_D^\infty\scalebox{.7 }[0.7 ]{\vdots}\nabla^0(\nabla{\textbf{\textit{U}}_0})\,,
 \label{GMRpre}
 \end{equation}
  with, following (\ref{comple1}) and (\ref{comple3}), $\hat{\textbf{\textit{F}}}_D^\infty\equiv\hat{\textbf{\textit{F}}}_D(\kappa\rightarrow0)=-6\pi\textbf{\textit{e}}_{\bf{3}}$ and $\hat{\textbf{\textit{S}}}_D^\infty\equiv\hat{\textbf{\textit{S}}}_D(\kappa\rightarrow0)=-2\pi{\textbf{\textit{I}}}\textbf{\textit{e}}_{\bf{3}}$. According to (\ref{carflow5}), $\nabla^0(\nabla\textbf{\textit{U}}_0)=(\nabla^2)^0\textbf{\textit{U}}_0=-2\alpha_c\textbf{\textit{e}}_{\bf{3}}$ is negative throughout the near-wall region and increases as the wall is approached through the rise of $\alpha_c$. Hence (\ref{GMRpre}) predicts that the slip velocity is negative (\textit{i.e.} the particle leads the fluid) and increases as the gap goes to zero. This is in total contradiction with the numerical profile displayed in figure \ref{Slip_1}$(a)$ which shows that, starting from zero far from the wall, the slip velocity becomes increasingly positive down to the wall.  \\
\indent Obviously the shortcoming of the GMR model in the present context is due to the omission of wall interaction effects. In the present theory, when the particle stands within the boundary layer,  the magnitude of these effects is influenced by the shape parameter $k_\delta$ involved in the approximate flow model (\ref{carflow5}). The discussion in \S\,\ref{BLM} suggested that the value $k_\delta=2$ properly describes the outer part of the boundary layer (where the particle stands when the separation distance is larger than $\Delta$, \textit{i.e.} $\epsilon>\Delta-1$), whereas $k_\delta=1$ much better describes the flow profile in the inner region relevant when $\epsilon\lesssim\Delta-1$. Figure \ref{Slip_1}$(b)$ shows the predictions of (\ref{Rezerod}) for the same small particle obtained with these two values of $k_\delta$; particles with a smaller or larger size behave similarly. First of all, it must be noticed that, unlike the GMR prediction in figure \ref{Slip_1}$(a)$, both predictions are in qualitative agreement with the numerical slip velocity profile. This emphasizes the crucial role of the repelling wall-induced Faxén force (\ref{faxen}) in the particle dynamics. Moreover, in line with the earlier discussion in \S\,\ref{BLM}, the figure confirms that the predicted profile obtained with $k_\delta=2$ agrees slightly better with numerical data for $\epsilon\gtrsim2$, while a much better agreement is obtained with $k_\delta=1$ for $\epsilon\lesssim1.5$. Hence the latter value is to be selected to obtain reliable predictions in the near-wall region, where the slip velocity exhibits large variations with the distance to the wall. \\
Last, figure \ref{Slip_prof0} compares predictions based on (\ref{Rezerod}) (with $k_\delta=1$) with numerical results for three different particle sizes corresponding to $\Delta^{-1}=0.3,\,0.4$ and $0.5$, \textit{i.e.} $\alpha Re=0.09,\,0.16$ and $0.25$, respectively. 
For each particle, the slip velocity is found to increase sharply as the particle approaches the wall. Moreover, the larger the particle the larger $\textbf{\textit{V}}_{S0}$ is when the dimensionless gap becomes small enough, typically $\epsilon\lesssim1.5$. These trends are well captured by the viscous prediction. However, (\ref{Rezerod}) starts to under-predict $\textbf{\textit{V}}_{S0}$ when the gap is such that $\epsilon\lesssim\Delta$. More precisely, for an increasing particle size, the viscous theory is found to underestimate the actual slip velocity at $\epsilon=1$ by $5\%$, $17\%$ and $22\%$, respectively. Therefore, the larger the particle, the stronger the under-estimate of $\textbf{\textit{V}}_{S0}$ is, a clear indication that inertial effects become responsible for an increasing fraction of the slip velocity as the particle size increases. At smaller gaps, the smallest particle displays a peculiar behaviour, since the slight underestimate observed for $\epsilon\gtrsim1$ almost vanishes. However, this agreement is presumably fortuitous since the asymptotic expressions involved in (\ref{Rezerod}) are barely accurate for such small gaps. We rather suspect that the corresponding simulation is slightly under-resolved in this case, owing to a marginally sufficient number of grid points per particle radius (see Appendix \ref{Simuli}).
\section{Leading-order inertial effects}
\label{inerteff}
\subsection{General considerations}
The above discussion sheds light on the limitations of the purely viscous force balance (\ref{Rezerod}) when the particle size increases and the wall is approached. To extend the validity of the theory toward larger particles, it is mandatory to include inertial corrections. Strictly speaking, only the limit of small-but-finite inertial effects can be tackled theoretically, which keeps the condition (\ref{cond3}) unchanged. Nevertheless, in practice one may hope the results of such a weakly-inertial theory to apply within an extended range of particle sizes satisfying the less restrictive condition $Re\ll\Delta^{-2}\lesssim1$. This is the goal of the developments summarized in the present section.\\
\indent The force balance (\ref{recipr}) is valid without any restriction regarding the magnitude of inertial effects. It provides the contribution of the velocity disturbance to these effects in the form of a volume integral over the entire flow domain. Examining the momentum equation for the disturbance under condition (\ref{cond1}) reveals that inertial terms become comparable to viscous terms at distances of $\mathcal{O}((\alpha Re)^{-1/2})$ from the particle. Hence, provided the latter is close enough to the wall for the condition 
\begin{equation}
\kappa^{-1}\lesssim(\alpha Re)^{-1/2}\iff\kappa^2\gtrsim \alpha Re
\label{cond2}
\end{equation}
to be satisfied, the flow field is properly approximated by the quasi-steady Stokes solution throughout the wall-particle gap. As recognized by \cite{Cox1968}, this in turn implies that in the outer region corresponding to distances $r\gtrsim(\alpha Re)^{-1/2}$ from the particle centre, the disturbance decays faster than in an unbounded domain, owing to the influence of the `image' field that cancels the disturbance at the wall. 
Because of this faster decay,  \cite{Cox1968} and \cite{Cox1977} showed that, within a large class of carrying flows, including the family of quadratic flows of interest here, the leading-order inertial corrections can be obtained through a regular perturbation procedure provided the particle is sufficiently close to the wall for (\ref{cond2}) to hold. Their argument was extended to unsteady situations in M1. Nevertheless, second-order inertial corrections of $\mathcal{O}(Re^2)$, $\mathcal{O}(\alpha Re^2)$ and $\mathcal{O}((\alpha Re)^2)$ remain associated with a singular perturbation, similar to the classical Oseen problem \citep{Proudman1957}. Therefore, a consistent description of small-but-finite inertial effects may be obtained solely \textit{via} a regular perturbation procedure only if the leading-order contributions are larger than the second-order ones. Provided (\ref{cond1}) holds, all the above second-order corrections are smaller than the $\mathcal{O}(\alpha Re)$-terms involved in the volume integral in the right-hand side of (\ref{recipr}). This is why we concentrate on the first three contributions to this volume integral in what follows.
\subsection{Effects of unsteadiness}
\label{unste}
\indent The inertial force associated with unsteady effects, namely ${\textbf{\textit{F}}}_U=-\alpha Re\int_\mathcal{V}(\hat{\textbf{\textit{U}}}+\textbf{\textit{e}}_{\bf{3}})\cdot(\partial\textbf{\textit{u}}/\partial t)d\mathcal{V}$ in (\ref{recipr}), was computed in M1 in the case where unsteadiness arises solely through time variations of the slip velocity. As far as $\alpha$ does not vary (\textit{i.e.} $\alpha_b=\alpha$ and $\alpha_c=0$ in (\ref{carflow5})) this contribution does not depend on the specific spatial structure of the carrying flow. Consequently, results derived in M1 apply directly to the present problem. In particular, equation (17b) of M1 provides the $\textbf{\textit{e}}_{\bf{3}}$-component of the unsteady contribution ${\textbf{\textit{F}}}_U$ in the form
\begin{equation}
{\textbf{\textit{F}}}_U\cdot\textbf{\textit{e}}_{\bf{3}}=-\frac{9}{4}\pi \alpha Re\left(\kappa^{-1}-\frac{13}{108}+\mathcal{O}(\kappa)\right)\frac{d\textbf{\textit{V}}_{S0}}{dt}\cdot\textbf{\textit{e}}_{\bf{3}}
\label{FU}
\end{equation}
This result only holds if the condition (\ref{cond2}) is satisfied, which makes the limit $\kappa\rightarrow0$ irrelevant. To understand the physical origin of this force, it is useful to evaluate its order of magnitude at the maximum wall-particle distance for which (\ref{FU}) is valid, \textit{i.e.} $\kappa^{-1}\sim(\alpha Re)^{-1/2}$. In this situation, the leading-order term in (\ref{FU}) is of $\mathcal{O}((\alpha Re)^{1/2})$. This is reminiscent of the magnitude of the `unsteady Oseen force' computed by \cite{Lovalenti1993} in the case of a particle with a finite slip Reynolds number accelerating or decelerating in an unbounded flow domain with the fluid at rest at infinity. Indeed, these authors found the unsteady Oseen force to be of $\mathcal{O}((St Re)^{1/2})$. Since $St\equiv \alpha$ here, the magnitude of ${\textbf{\textit{F}}}_U$ predicted by (\ref{FU}) for $\kappa^{-1}\sim(\alpha Re)^{-1/2}$ is similar to that of the inertial force they computed. This is a strong indication that $\textbf{\textit{F}}_U$ is not a force that originates from the wall, but is merely what is left from the unsteady Oseen force as the wall is approached. Starting from a magnitude of $\mathcal{O}((\alpha Re)^{1/2})$ for large separation distances ($\kappa\rightarrow0$), the unsteady Oseen force is gradually weakened by the wall as $\kappa$ increases and becomes of $\mathcal{O}(\alpha Re)$ for small separations ($\kappa\rightarrow1$). The prediction (\ref{FU}) expresses this near-wall variation for moderate-to-small separation distances such that $\kappa\gtrsim(\alpha Re)^{1/2}$. \cite{Lovalenti1993} showed that the unsteady Oseen force primarily results from the time variations of the wake structure due to the particle acceleration or deceleration. 
Any disturbance originating in a time variation of $\textbf{\textit{V}}_{S0}$ requires a finite time to diffuse away from the particle surface and reach the wake region. For this reason, the expression for this force in the case of an unbounded fluid domain involves a convolution integral. The corresponding kernel, inertial by nature, is distinct from that associated with the Basset-Boussinesq force, which originates in the unsteady diffusion of vorticity close to the particle. The near-wall situation considered here, combined with the slow evolution implied by the restriction $ReSt\equiv \alpha Re\ll1$, drastically reduces the above finite memory effect. Indeed, these slow variations imply that the leading-order contribution to the disturbance $\textit{\textbf{u}}$ is governed by the quasi-steady Stokes equation at distances less than $(ReSt)^{-1/2}$. Since the dominant contribution to the near-wall unsteady effects is provided by a regular perturbation procedure, only this quasi-steady disturbance is involved, making the resulting force only dependent on the current acceleration $d\textbf{\textit{V}}_{S0}/dt$. The same happens with the contribution due to the time rate-of-change of the near-particle disturbance, which usually yields the Basset-Boussinesq force and is here also encapsulated in the $\mathcal{O}(\kappa^{-1})$-term of (\ref{FU}), while the added-mass contribution and second-order corrections associated with the unsteady Oseen force form the $\mathcal{O}(\kappa^{0})$-term. Hence the entire contribution of unsteady effects at any time is expressible solely in terms of the current acceleration $d\textbf{\textit{V}}_{S0}/dt$ when the particle gets close enough to the wall and time variations are slow enough for the condition $\alpha Re\ll1$ to be satisfied.
 Note that, since $\kappa$ is less than 1 by definition, the $\kappa^{-1}$-term is always dominant in (\ref{FU}). Hence ${\textbf{\textit{F}}}_U$ always tends to lower the relative acceleration $d\textbf{\textit{V}}_{S0}/dt$, just as the familiar added-mass effect does. \vspace{2mm}\\
 \indent When the particle stands within the boundary layer, other sources of unsteadiness arise through the time-dependent strain rates $\alpha_b(t)$ and $\alpha_c(t)$. Since $d\kappa/dt=-\alpha^{-1}\kappa^2\textbf{\textit{V}}\cdot\textbf{\textit{e}}_{\bf{3}}$, the definitions of $\alpha_b$ and $\alpha_c$ in (\ref{coeff}) imply that $d\alpha_b/dt=2\kappa\frac{\Lambda^2}{(1+\Lambda)^3}\textbf{\textit{V}}\cdot\textbf{\textit{e}}_{\bf{3}}$ and $d\alpha_c/dt=-6\kappa^2\frac{\Lambda^2}{(1+\Lambda)^4}\textbf{\textit{V}}\cdot\textbf{\textit{e}}_{\bf{3}}$. To express the corresponding contributions to the force, it is convenient to split the particle velocity in the form $\textbf{\textit{V}}=\textbf{\textit{V}}_{S0}+\textbf{\textit{U}}_0^0$, with $\textbf{\textit{U}}_0^0$ as given in (\ref{coeff}). Keeping in mind that $\alpha Re\Lambda^2=k_\delta^{-2}\kappa^2$ and that $\Lambda=\mathcal{O}(1)$ for $\kappa=\mathcal{O}(\Delta^{-1})$, variations of $\alpha_b(t)$ are found to contribute to generate a nonzero slip through an $\mathcal{O}(\kappa^2)$-source term (since $\textbf{\textit{U}}_0^0\propto\kappa^{-1}$), and an $\mathcal{O}(\kappa^3)$-correction to the pre-factor of the force contribution proportional to $\textbf{\textit{V}}_{S0}$, \textit{i.e.} to the drag coefficient. Variations of $\alpha_c(t)$ provide contributions smaller by an $\mathcal{O}(\frac{\kappa}{1+\Lambda})$-factor. Let us first consider the force resulting from $\alpha_b(t)$-variations. The procedure employed to compute this contribution and all those to come in this section is summarized in Appendix \ref{CIE}. According to (\ref{FI20}) and the considerations that follow, this force is found to be
\begin{eqnarray}
\nonumber
 {\textbf{\textit{F}}}_{U\delta}\cdot\textbf{\textit{e}}_{\bf{3}}&\approx&-\frac{15}{4}\pi\alpha Re\frac{d\alpha_b}{dt}(1+\frac{9}{8}\kappa)(\textbf{\textit{U}}_0^0+\textbf{\textit{V}}_{S0})\cdot\textbf{\textit{e}}_{\bf{3}}\\
 &\approx&\frac{15\pi}{2k_\delta^2}\frac{\kappa^2}{(1+\Lambda)^3}\left\{\frac{2\alpha}{1+\Lambda}(1+\frac{9}{8}\kappa)-\kappa\textbf{\textit{V}}_{S0}\cdot\textbf{\textit{e}}_{\bf{3}}\right\}\,,
 \label{FI22}
 \end{eqnarray}
 where the second approximation is obtained by incorporating the explicit expressions for $d\alpha_b/dt$ and $\textbf{\textit{U}}_0^0$.
 The source term in (\ref{FI22}) (first term within braces) is positive, contributing to make the particle lag behind the fluid. That a body translating steadily perpendicular to a wall generates a nonzero normal force directly through the time variation of its position is not uncommon. In particular, this is the case in the inviscid limit, where the increase of the fluid volume entrained by the body as it gets closer to the wall results in a repulsive force, just as in (\ref{FI22}) \citep{MilneThomson1962}. \\
\indent  At this point it is useful to compare the magnitude of the $\mathcal{O}(\kappa^3)$-terms in (\ref{FI22}) with those involved in the zero-$Re$ approximation (\ref{Rezerod}), keeping in mind that $\Lambda$ becomes large when $\kappa\rightarrow1$. To fix ideas, let us consider a particle $10$ times smaller than the boundary layer thickness, \textit{i.e.} $\Delta=10$, standing at the position corresponding to $\kappa=1/2$. With $k_\delta=1$ one then has $\Lambda=5$. Consequently the ratio of the $\mathcal{O}(\kappa^3)$-source term in (\ref{FI22}) to its counterpart in the curvature-induced Faxén term in (\ref{Rezerod}) is of $\mathcal{O}(\frac{1}{\Lambda^2(1+\Lambda)})\approx0.007$. Similarly, the ratio of the $\mathcal{O}(\kappa^3)$-drag correction in (\ref{FI22}) (second term within braces) to the corresponding term in (\ref{Rezerod}) is of $\mathcal{O}(\frac{1}{(1+\Lambda)^3})\approx0.005$. These estimates indicate that  $\mathcal{O}(\kappa^3)$-corrections weighted by a $\frac{1}{(1+\Lambda)^n}$-factor with $n\ge3$ are negligibly small at the present order of approximation. For this reason, such terms will be systematically dropped in what follows, and only the leading-order $\mathcal{O}(\kappa^2)$-source term present in (\ref{FI22}) will be conserved when $ {\textbf{\textit{F}}}_{U\delta}$ will be inserted in the final force balance. As mentioned above, contributions involved in the force correction resulting from variations of $\alpha_c(t)$ are smaller than those induced by $\alpha_b(t)$-variations by an $\mathcal{O}(\frac{\kappa}{1+\Lambda})$-factor. Hence the previous argument shows that all of them are negligible at the present order of approximation. For the  same reason, the last two terms within parentheses in the right-hand side of (\ref{F0BL}) also provide a negligible contribution to the inertial force $\textbf{\textit{F}}_0$.


\subsection{Effects of advective transport}
\label{adve}
\indent Within the framework of the above conditions, especially (\ref{cond1}), the other contributions to be considered in the volume integral of the right-hand side of (\ref{recipr}) are the advective terms proportional to $\alpha Re$, which result from the quasilinear contribution $\textbf{\textit{u}}\cdot\nabla{\textbf{\textit{U}}}_0(\textbf{\textit{x}},t)+(\textbf{\textit{U}}_0(\textbf{\textit{x}},t)-\textbf{\textit{U}}_0^0(t))\cdot\nabla\textbf{\textit{u}}$ in the disturbance momentum equation. Due to the ambient strain, the leading-order contribution to the disturbance arises from a stresslet. For this reason, its advective transport by the linear flow component (and \textit{vice versa}) yields a contribution of $\mathcal{O}(\alpha_b^2Re)$.
\cite{Cox1977} evaluated a similar term in the case of a uniformly sheared carrying flow, where it yields a net lift force on the particle; their prediction was later confirmed by \cite{Cherukat1994}. Although the scaling of this force with respect to $\alpha_b$, $Re$ and $\kappa$ does not depend on the specific linear base flow under consideration, the pre-factor that determines its actual strength does. To the best of our knowledge, this contribution, say $ {\textbf{\textit{F}}}_{I}$, has not been evaluated so far in the axisymmetric straining flow (\ref{carflow}).
   Based on (\ref{FI10}) and the considerations that follow, the final result valid up to $\mathcal{O}(\kappa)$ is
  \begin{equation}
 {\textbf{\textit{F}}}_{I}\cdot\textbf{\textit{e}}_{\bf{3}}\approx\left(1+\frac{9}{8}\kappa\right)\frac{75}{16}\pi\alpha_b^2 Re\,.
 \label{FI}
 \end{equation}
 The force ${\textbf{\textit{F}}}_{I}$ arises due to the asymmetry created by the wall in the transport of the stresslet by the straining flow and \textit{vice versa}.  In a linear shear flow, the counterpart of ${\textbf{\textit{F}}}_{I}$ involves a pre-factor $\frac{55}{96}\pi$ instead of $\frac{75}{16}\pi$ \citep{Cox1977}. Consequently, the magnitude of ${\textbf{\textit{F}}}_{I}$ is approximately $8.2$ times larger in the present axisymmetric straining flow than in a uniform shear with strength $\alpha_b$. Similar to that of ${\textbf{\textit{F}}}_U$, the above prediction for ${\textbf{\textit{F}}}_{I}$ only holds up to a maximum separation of $\mathcal{O}((\alpha_b Re)^{-1/2})$. For larger separations, ${\textbf{\textit{F}}}_{I}$ must tend to zero as $\kappa\rightarrow0$ but this decay cannot be captured by the regular expansion procedure employed here.\\
 \indent Within the boundary layer, several additional contributions arise, due to the presence of the quadratic flow component in (\ref{carflow5}). A detailed examination of their respective magnitudes reveals that the largest one is provided by the transport of the leading $\mathcal{O}(\alpha_b)$-stresslet by the quadratic $\mathcal{O}(\alpha_c)$-flow component and \textit{vice versa}. This mechanism results in an $\mathcal{O}(\kappa^{-1}\alpha_b\alpha_cRe)$-force, the formal expression of which takes the form (\ref{FI3}). As outlined in Appendix \ref{CIE}, numerical evaluation of this expression and truncation considerations based on the argument discussed at the end of \S\,\ref{unste} lead to
 \begin{equation}
 {\textbf{\textit{F}}}_{I\delta}\cdot\textbf{\textit{e}}_{\bf{3}}\approx\frac{85}{8}\alpha_b\alpha_cRe\kappa^{-1}=\frac{85}{4}\alpha\frac{1+2\Lambda}{(1+\Lambda)^5}\kappa^2\,,
  \label{FI32}
 \end{equation}
where the last equality results from the definitions of $\alpha_b$ and $\alpha_c$ in (\ref{coeff}) and the relation $\Lambda^2=\kappa^2/(k_\delta^2\alpha Re)$.\vspace{2mm}\\
 \indent Another inertial effect results from the transport of the Stokeslet associated with the slip velocity by the ambient straining flow and \textit{vice versa}. This advective process yields a force whose leading-order contribution is proportional to $\alpha_b Re\kappa^{-1}{\textbf{\textit{V}}}_{S0}$. Since the zero-$Re$ force balance (\ref{Rezerod}) suggests that the slip velocity is of $\mathcal{O}(\kappa^2\alpha_b)$, this force correction is expected to be of $\mathcal{O}(\kappa\alpha_b^2Re)$, \textit{i.e.} smaller than ${\textbf{\textit{F}}}_I$ by an $\mathcal{O}(\kappa)$-order of magnitude. Nevertheless, for $\kappa=\mathcal{O}(\alpha_b Re)^{1/2}$, $\alpha_b Re\kappa^{-1}{\textbf{\textit{V}}}_{S0}=(\alpha_b Re)^{1/2}{\textbf{\textit{V}}}_{S0}$. Hence this effect provides a correction to the drag which is for instance larger than the second term in the inertial force ${\textbf{\textit{F}}}_0$ in (\ref{F0}) and must be included for consistency. Details regarding the computation of this contribution are also provided in Appendix \ref{CIE} (see (\ref{FDA0}) and the comments that follow). Its final expression is found to be 
  \begin{equation}
 {\textbf{\textit{F}}}_{D\alpha}\cdot\textbf{\textit{e}}_{\bf{3}}\approx\frac{\pi}{16}\alpha_b Re\left(45\kappa^{-1}-\frac{1861}{60}\right)\textbf{\textit{V}}_{S0}\cdot\textbf{\textit{e}}_{\bf{3}}\,.
 \label{FDA}
 \end{equation}
 For similar reasons, the contribution resulting from the transport of the Stokeslet associated with the slip velocity by the quadratic flow and \textit{vice versa} must also be considered. The corresponding force is proportional to $\alpha_c Re\kappa^{-2}{\textbf{\textit{V}}}_{S0}\sim\frac{\kappa}{k_\delta^2(1+\Lambda)^3}{\textbf{\textit{V}}}_{S0}$. As outlined in Appendix \ref{CIE}, evaluating the corresponding volume integral and  truncating the result in line with the discussion in \S\,\ref{unste} yields 
\begin{equation}
{\textbf{\textit{F}}}_{D\alpha\delta} \cdot\textbf{\textit{e}}_{\bf{3}}\approx\frac{3}{4}\pi\alpha_cRe\kappa^{-2}(1+\frac{9}{4}\kappa)\textbf{\textit{V}}_{S0}\cdot\textbf{\textit{e}}_{\bf{3}}=\frac{3}{2}\pi\frac{\kappa}{k_\delta^2(1+\Lambda)^3}(1+\frac{9}{4}\kappa)\textbf{\textit{V}}_{S0}\cdot\textbf{\textit{e}}_{\bf{3}}\,.
\label{FDC}
\end{equation}
It is worth noting that, although the boundary-layer contributions (\ref{FI22}), (\ref{FI32}) and (\ref{FDC}) are inertial by nature, the strain Reynolds number $\alpha Re$ no longer appears in their final expression once $\textbf{\textit{U}}_{0}^0$, $\alpha_b$ and $\alpha_c$ have been replaced by their definitions as given in (\ref{coeff}). This is because they are proportional to $d\alpha_b/dt$ or $\alpha_c$, both of which are proportional to $\kappa\Lambda^2\propto\kappa^3\Delta^2$, and $\Delta$ equals $(\alpha Re)^{-1/2}$. 
 \subsection{Final force balance}
 \label{FFB}
All contributions computed in \S\S\,\ref{unste} and \ref{adve} may finally be gathered to enhance (\ref{Rezerod}) with effects of finite fluid inertia. The updated force balance can be expressed in the form 
 \begin{equation}
 ({\textbf{\textit{F}}}_0-{\textbf{\textit{F}}}_U)\cdot{\bf{e}_3}-\hat{\textbf{\textit{F}}}_D\cdot\textbf{\textit{V}}_{S0}-(\textbf{\textit{F}}_{D\alpha}+\textbf{\textit{F}}_{D\alpha\delta})\cdot\textbf{\textit{e}}_{\bf{3}}=({\textbf{\textit{F}}}_{F}+{\textbf{\textit{F}}}_{F\delta}+{\textbf{\textit{F}}}_I+{\textbf{\textit{F}}}_{I\delta}+{\textbf{\textit{F}}}_{U\delta})\cdot\textbf{\textit{e}}_{\bf{3}}\,,
 \label{balance}
 \vspace{2mm}
 \end{equation}
with ${\textbf{\textit{F}}}_0$ as given in (\ref{F0BL}), and ${\textbf{\textit{F}}}_U$, ${\textbf{\textit{F}}}_{U\delta}$, ${\textbf{\textit{F}}}_I$, ${\textbf{\textit{F}}}_{I\delta}$, $\textbf{\textit{F}}_{D\alpha}$ and $\textbf{\textit{F}}_{D\alpha\delta}$ taken from (\ref{FU})-(\ref{FDC}).
We then define the ratios
\begin{equation}
A_\Lambda=\frac{1+2\Lambda}{(1+\Lambda)^2}\,,\quad B_\Lambda=\frac{1}{k_\delta^2(1+\Lambda)^3}\,,\quad C_\Lambda=\frac{1}{k_\delta^2(1+\Lambda)^4}\,,\quad D_\Lambda=\frac{\Lambda^2}{(1+\Lambda)^3}\,.
\label{definl}
\end{equation}
Boundary layer effects become negligible in the double limit $\Lambda\ll1$ (\textit{i.e.} the separation $\kappa^{-1}$ is very large compared to $\Delta/k_\delta$) and $k_\delta\rightarrow\infty$ (\textit{i.e.} the fluid layer within which the no-slip condition at the wall significantly influences the carrying flow is much thinner than $\Delta$)
, in which case $A_\Lambda\rightarrow1$ and $B_\Lambda$, $C_\Lambda$ and $D_\Lambda\rightarrow0$. Nevertheless, condition (\ref{cond2}) implies that the inertial corrections derived in \S\S\, \ref{unste} and \ref{adve} are valid only for $\kappa\gtrsim\Delta^{-1}$, \textit{i.e.} $\Lambda\gtrsim k_\delta^{-1}$. Therefore predictions involving these corrections are not expected to be relevant for small values of $\Lambda$. As already mentioned, $\Lambda$ is large when $\kappa\rightarrow1$ since we are considering small particles. Consequently all four ratios in (\ref{definl}) go through $\mathcal{O}(1)$-values in some intermediate range of $\kappa$ and become small in the limit $\kappa\rightarrow1$. 
The final approximate force balance (\ref{balance}) takes the form
 \begin{eqnarray}
 \label{Refini2}
&&9\alpha Re\left(\kappa^{-1}+\frac{17}{36}\right)\frac{d\textbf{\textit{V}}_{S0}}{dt}\cdot\textbf{\textit{e}}_{\bf{3}}\\
\nonumber
&&+24\Bigg\{1+\frac{9}{8}\kappa+\frac{81}{64}\kappa^2+\frac{473}{512}\kappa^3-\frac{1}{4}\kappa \left(1+\frac{9}{4}\kappa\right)B_\Lambda
-\frac{15}{32}\alpha Re\left(\kappa^{-1}+\frac{4421}{2700}\right)A_\Lambda\Bigg\}\textbf{\textit{V}}_{S0}\cdot\textbf{\textit{e}}_{\bf{3}}\\
\nonumber
&&\approx\alpha \bigg\{\kappa^2A_\Lambda\left(45(1+\frac{9}{8}\kappa)+85B_\Lambda\right)+60\kappa^2C_\Lambda-16\kappa\left(1+\frac{9}{8}\kappa+\frac{81}{64}\kappa^2\right) D_\Lambda+\frac{75}{4}\alpha Re(1+\frac{9}{8}\kappa)A_\Lambda^2\bigg\}\,.
\end{eqnarray}
\\
 Inertial forces ${\textbf{\textit{F}}}_{I}$ and ${\textbf{\textit{F}}}_{I\delta}$ resulting from the advective transport of the stresslet by the linear and quadratic flow components, respectively, and ${\textbf{\textit{F}}}_{U\delta}$ resulting from the time-variation of the straining rate about the particle, all provide positive contributions to the right-hand side of (\ref{Refini2}). Hence they all contribute to make the particle lag behind the fluid (since $\textbf{\textit{U}}_{0}^0\cdot\textbf{\textit{e}}_{\bf{3}}<0$), similar to the wall-induced Faxén force ${\textbf{\textit{F}}}_{F}$. Only the curvature-induced Faxén force ${\textbf{\textit{F}}}_{F\delta}$ tends to make the particle lead the fluid; the smaller the particle the larger the relative influence of this force at a given distance from the wall. Consider for instance a particle standing a distance $\Delta$ from the wall, \textit{i.e.} $\Lambda=1$. The right-hand side of (\ref{Refini2}) then becomes negative only if ${\Delta}^{-1}\lesssim0.037$. Comparing with the prediction provided by the zero-$Re$ approximation (\ref{Rezerod}) indicates that inertial effects lower the critical size of particles for which the driving force changes sign at this location by a factor of $1.6$. Alternatively, inertial effects may be said to enhance the positive slip between the particle and the fluid. Moreover, all inertial terms that contribute to the $B_\Lambda$- and $\alpha ReA_\Lambda$-terms in the pre-factor of the $\textbf{\textit{V}}_{S0}$-term, namely forces ${\textbf{\textit{F}}}_{D\alpha}$ and ${\textbf{\textit{F}}}_{D\delta}$ resulting from the transport of the Stokeslet by the linear and quadratic flow components, respectively, and the advective part of the force ${\textbf{\textit{F}}}_{0}$ due to the acceleration of the undisturbed flow, decrease the drag coefficient. Hence they all tend to enhance the slip velocity for a given value of the overall source term, reinforcing the role of inertia in the slip increase. Incidentally, this points out to the fact that, unlike the usual inertial increase of the drag coefficient encountered in the classical Oseen problem \citep{Proudman1957}, inertial corrections in the HH flow lower the drag coefficient.
\subsection{Comparison with numerical results}
\label{compan}
\begin{figure}
\centering
\vspace{2mm}
\begin{subfigure}[]
\centering
\includegraphics[width=0.492\textwidth]{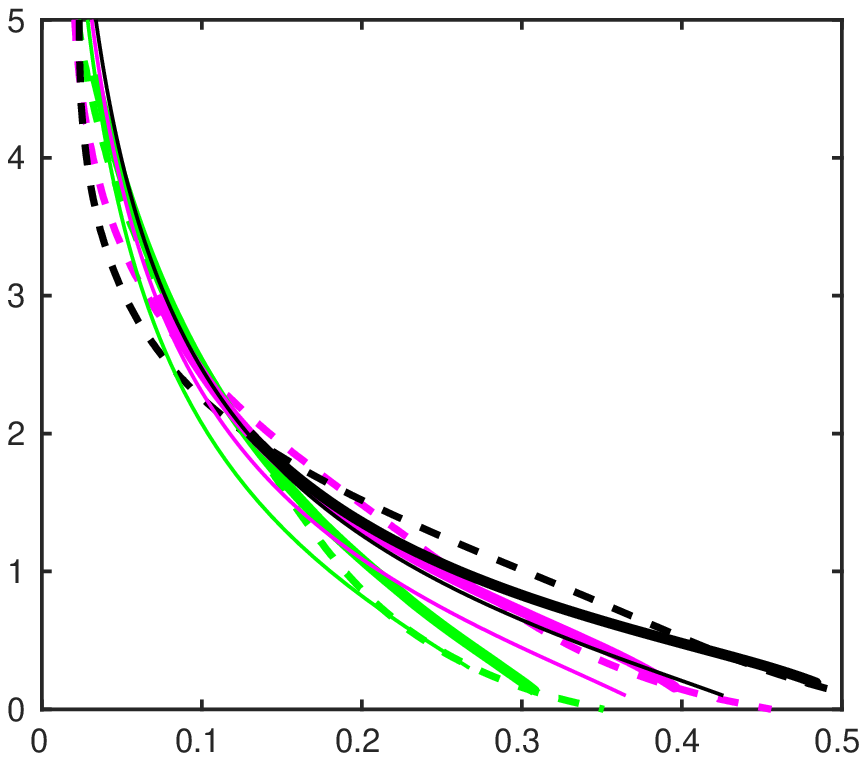}
    \end{subfigure} 
    \hfill
\begin{subfigure}[]
\centering
\includegraphics[width=0.492\textwidth]{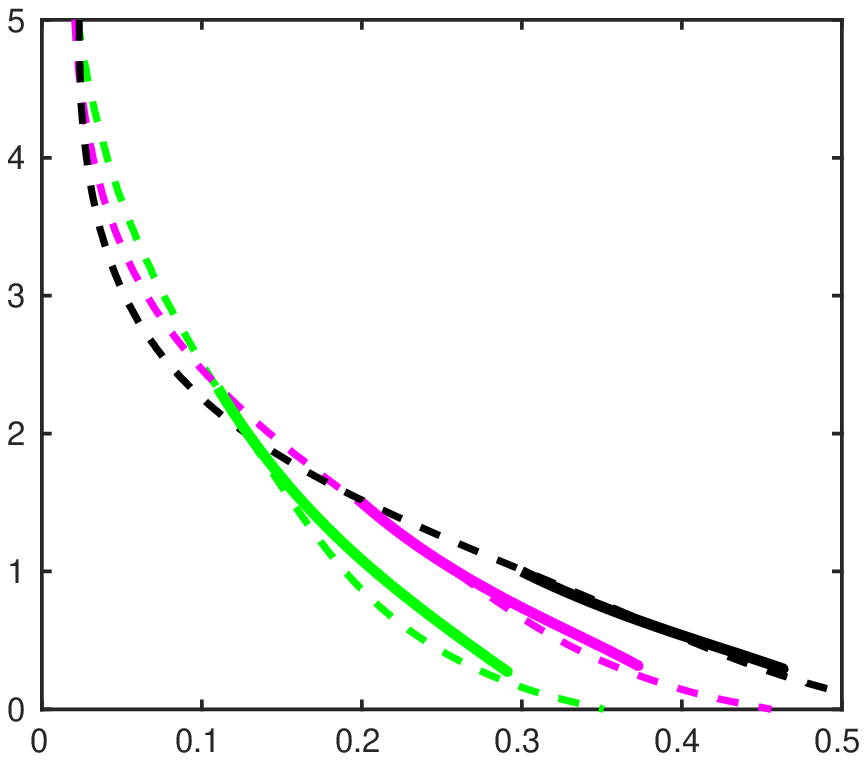}
    \end{subfigure}
    \hfill
\begin{flushleft}
\vspace{-46.5mm}\hspace{4mm}$\epsilon$\hspace{67mm}$\epsilon$\\
\vspace{33.8mm}
\hspace{25mm}$\alpha^{-1}\textbf{\textit{V}}_{S0}\cdot\textbf{\textit{e}}_{\bf{3}}$\hspace{50.5mm}$\alpha^{-1}\textbf{\textit{V}}_{S0}\cdot\textbf{\textit{e}}_{\bf{3}}$
\end{flushleft}
\vspace{0mm}
\caption{Predictions for the slip velocity in the near-wall region for particles with increasing relative size $\Delta^{-1}=0.3$ (\protect\greenline), $0.4$ (\protect\magentaline), $0.5$ (\protect\blackline); all predictions are based on the undisturbed flow (\ref{carflow4}) with $k_\delta=1$. $(a)$: initial gap $\epsilon_i=3\Delta/2-1$; $(b)$: $\epsilon_i=\Delta-1$.  
Dashed line: simulation results; thick solid line: finite-$Re$ prediction from (\ref{Refini2}); thin solid line in $(a)$: creeping-flow prediction (\ref{Rezerod}) (the thick purple line and the thin black line almost overlap). 
}
\label{Slip_prof}
\end{figure}
Unlike the purely viscous solution (\ref{Rezerod}), predictions involving inertial corrections are only meaningful within a limited separation range, since (\ref{Refini2}) is expected to be valid only in the near-wall region such that $\kappa\gtrsim\Delta^{-1}$. Consequently, the larger the particle the smaller the separation range over which the comparison between predictions of (\ref{Refini2}) and results of fully-resolved simulations is relevant. 
As (\ref{Refini2}) is a first-order differential equation with respect to $\textbf{\textit{V}}_{S0}$, an initial condition for the slip velocity is required. If the expressions obtained for the inertial corrections were valid up to large separations, $\textbf{\textit{V}}_{S0}(t=0)=\textbf{0}$ in the limit $\kappa\rightarrow0$ would be a natural choice. Given their limited range of validity, an alternative is required. Without results from fully-resolved simulations available, the most obvious choice is to use the slip velocity provided by the viscous prediction (\ref{Rezerod}) to initialize the determination of $\textbf{\textit{V}}_{S0}$ at a position $\kappa_i$ such that $\kappa_i=\mathcal{O}((\alpha Re)^{1/2})=\mathcal{O}(\Delta^{-1})$. Since figure \ref{Hiemenz_DNS} indicates that the carrying flow model (\ref{carflow5}) correctly fits the actual HH profile with $k_\delta=1$ up to a distance to the wall of approximately $1.5\Delta$, we select $\kappa_i=(1.5\Delta)^{-1}$, \textit{i.e.} $\epsilon_i=3\Delta/2-1$, a position at which the creeping-flow approximation (\ref{Rezerod}) and the fully-resolved simulation predict close values of the slip velocity. 
Based on this initialization protocol, figure \ref{Slip_prof}$(a)$ compares predictions of (\ref{Refini2}) with simulation results for the three particles already considered in figure \ref{Slip_prof0}. 
In all cases, inertial effects are seen to increase the slip velocity at a given separation distance (compare the predictions corresponding to the thin and thick solid lines for each particle). This is because all inertial terms in the right-hand side of (\ref{Refini2}) are positive, while all inertial corrections to the drag coefficient in the left-hand side are negative. Moreover, since $\alpha Re=\Delta^{-2}$ and all coefficients $A_\Lambda-D_\Lambda$ are decreasing functions of $\Lambda$ (hence of $\Delta$), increasing the particle size, \textit{i.e.} $\Delta^{-1}$, makes all inertial terms in the right-hand side increase at a given $\kappa$. Because of this, the larger the particle the stronger the inertial correction to the slip at a given distance from the wall is. Both features act to compensate for the deficiencies of the purely viscous force balance (\ref{Rezerod}) analyzed in \S\,\ref{compaDNS}. This makes the weakly-inertial prediction based on (\ref{Refini2}) significantly closer to the numerical solution for moderate-to-small gaps (the agreement deteriorates at small gaps for the smallest particle, owing to the peculiar behaviour of the numerical prediction mentioned in \S\,\ref{compaDNS}).\\
Nevertheless, a closer look at the slip velocity profiles in figure \ref{Slip_prof}$(a)$ shows that the slope $d(\textbf{\textit{V}}_{S0}\cdot\textbf{\textit{e}}_{\bf{3}})/d\kappa$ is underestimated for $\epsilon\lesssim\epsilon_i$, which maintains the predicted values of $\textbf{\textit{V}}_{S0}$  slightly below those found in the simulations down to $\epsilon\approx\epsilon_i/3$. To get some insight into the origin of this shortcoming, it is of interest to consider the predictions of (\ref{Refini2}) obtained by selecting a smaller initial separation, $\kappa_i=\Delta^{-1}$, \textit{i.e.} $\epsilon_i=\Delta-1$. Since the viscous force balance (\ref{Rezerod}) significantly underestimates the actual slip velocity at this smaller separation (see figure \ref{Slip_prof0}), we employed the value $\textbf{\textit{V}}_{S0}(\epsilon_i=\Delta-1)$ provided by the fully-resolved simulations as initial condition in this case. As figure \ref{Slip_prof}$(b)$ shows, the prediction resulting from (\ref{Refini2}) now closely agrees with the simulation results for $\epsilon\le\epsilon_i$, especially for the largest two particles. The agreement extends down to a dimensionless gap $\epsilon\approx0.3$ $(\kappa\approx3/4)$, significantly beyond the expected limit of validity ($\epsilon\approx1$) of the truncated asymptotic expression of the `auxiliary' solution. The reason why the slope $d(\textbf{\textit{V}}_{S0}\cdot\textbf{\textit{e}}_{\bf{3}})/d\kappa$ is correctly predicted when $\epsilon_i=\Delta-1$ but is underestimated when $\epsilon_i=3\Delta/2-1$ is readily identified in (\ref{Refini2}), keeping in mind that the term that absorbs the local variations of $\textbf{\textit{V}}_{S0}$ is the unsteady force proportional to $d\textbf{\textit{V}}_{S0}/dt$. As discussed in \S\,\ref{unste}, the expression (\ref{FU}) for this contribution is dominated by a term proportional to $\kappa^{-1}$. The growth of this term with the separation distance is only correct as far as the wall stands in the inner region of the disturbance. For larger separations, it becomes unphysical, since the entire contribution must tend toward the finite `unsteady Oseen force' computed by \cite{Lovalenti1993} when $\kappa\rightarrow0$. This unphysical growth makes this force overestimated for $\kappa\lesssim(\alpha Re)^{1/2}$ and is responsible for the slight underestimate of $\textbf{\textit{V}}_{S0}$ noticed for $\epsilon\lesssim\epsilon_i$ in figure \ref{Slip_prof}$(a)$. 
This analysis leads to the conclusion that the technical bottleneck that restricts most the validity of (\ref{Refini2}) towards larger separations is the limited range of validity of (\ref{FU}). This calls for a specific study aimed at deriving the proper expression for the unsteady Oseen force in the case where the particle is already influenced by the wall but the latter stands in the outer region of the disturbance.

\section{A particle released off-axis}
\label{Off}
\subsection{Preliminaries}
\label{prelim}
\begin{figure}
\vspace{-55 mm}
\hspace{8 mm}{\includegraphics[width=0.99\textwidth]{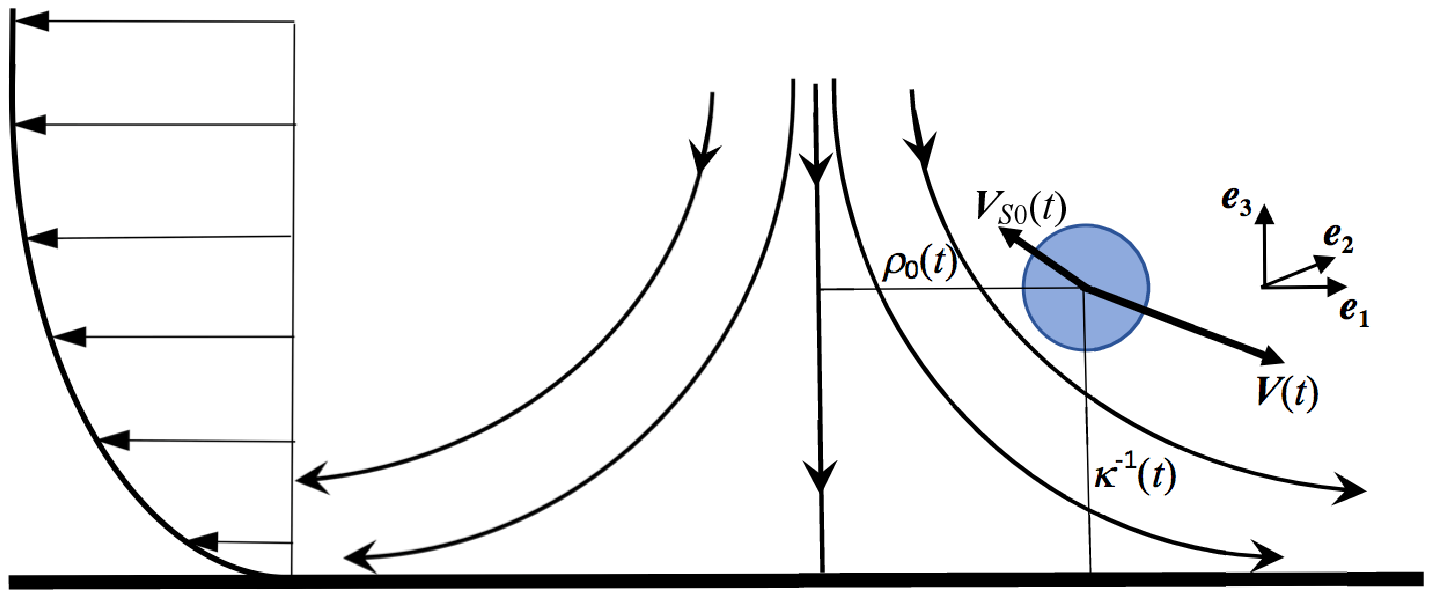}}
\vspace{-88 mm}
\caption[ ]{\footnotesize{Sketch of the configuration with the particle released some distance from the axis of the HH flow.}}
\label{sketch_off}
\end{figure}

Up to now, we constrained the particle to move along the symmetry axis of the HH flow. Although the simulations of \cite{Li2019} only addressed this case, it represents a quite specific situation. The techniques used to obtain the various wall-normal forces in \S\S\,\ref{ZeroRe} and \ref{inerteff} may also be applied to predict the wall-parallel slip velocity component and the modifications of the slip wall-normal component when the particle stands an arbitrary time-dependent radial distance from the axis, say $\rho_0(t)$, as sketched in figure \ref{sketch_off} 
 In order for the flow to satisfy the no-slip boundary condition at the wall whatever $\textbf{\textit{x}}_{0\parallel}=\rho_0(t)\textbf{\textit{e}}_{\bf{1}}$, the radial position $\textbf{\textit{x}}_\parallel$ involved in (\ref{carflow4}) has to be changed into $\textbf{\textit{x}}_\parallel+\textbf{\textit{x}}_{0\parallel}$ (hence $\textbf{\textit{x}}$ into $\textbf{\textit{x}}+\textbf{\textit{x}}_{0\parallel}$). With this transformation, the undisturbed flow field in the vicinity of the particle ($|x_3| \ll(1+\Lambda)/\kappa$) takes the form 
\begin{equation}
\textbf{\textit{U}}_0(\textbf{\textit{x}},t)=\textbf{\textit{U}}_{0}^{\rho_0}(t)+\{\alpha_b(t)({\textbf{\textit{x}}}-3x_3\textbf{\textit{e}}_{\bf{3}})+\alpha_c(t)x_3({\textbf{\textit{x}}}-2x_3\textbf{\textit{e}}_{\bf{3}})\}+\rho_0(t)\{\alpha_c(t)x_3+\alpha_d(t)x_3^2\}\textbf{\textit{e}}_{\bf{1}}+...\,,
\label{carflow5p}
\end{equation}
with 
\begin{equation}
\textbf{\textit{U}}_{0}^{\rho_0}(t)=\textbf{\textit{U}}_{0}^{0}(t)+\alpha_b\rho_0\textbf{\textit{e}}_{\bf{1}}\quad \mbox{and}\quad \alpha_d(t)=-3\alpha\kappa^2\frac{\Lambda^2}{(1+\Lambda)^4}\,,
\label{coeffp}
\end{equation}
$\alpha_b(t)$, $\alpha_c(t)$ and $\textbf{\textit{U}}_{0}^{0}(t)$ being still as given in (\ref{coeff}). Compared to (\ref{carflow5}), (\ref{carflow5p}) reveals that, at a radial position $\rho_0$ from the axis, the undisturbed flow  comprises an additional shear component proportional to $\rho_0(t)\alpha_c(t)$, a parabolic component proportional to $\rho_0(t)\alpha_d(t)$ \textit{etc.}, all of which correspond to a radial flow whose intensity increases linearly with $\rho_0$. As time elapses, the particle is transported away from the axis $\rho_0=0$ by the carrying flow. Therefore $\rho_0(t)$ increases, which makes the radial component in (\ref{carflow5p}) increase at the expanse of the axial wall-normal component. In other words, the flow in the vicinity of the particle looks more and more like a wall-parallel shear flow. \\
\indent Let us provisionally consider that the particle stands beyond the boundary layer. 
Compared to the axisymmetric configuration contemplated so far, there is no change in the strain-induced disturbance, since the straining motion is identical to that in (\ref{carflow}). In particular, the disturbance does not depend on the radial position  $\rho_0$. Consequently, all forces which only depend on the strain rate and the distance to the wall are unchanged. This remark enables us to conclude that no source term for the parallel slip component can exist as far as the particle has not entered the boundary layer, even though inertial effects are taken into account. Indeed, the two contributions $\textbf{\textit{F}}_F$ in (\ref{faxen}) and $\textbf{\textit{F}}_I$ in (\ref{FI}) result from the interaction of the $\rho_0$-independent stresslet with the wall, so that any nonzero $\textbf{\textit{e}}_{\bf{1}}$-component of one of these forces would be $\rho_0$-independent. Since no radial force component can exist when the particle stands on the flow axis, such a component remains null whatever $\rho_0$. \\
\indent To obtain the various contributions to the radial force within the boundary layer, we need to project the reciprocal theorem onto the $\textbf{\textit{e}}_{\bf{1}}$-direction. The result is similar to (\ref{recipr}), except that the unit vector $\textbf{\textit{e}}_{\bf{3}}$ has to be replaced with $\textbf{\textit{e}}_{\bf{1}}$ everywhere, and the relevant auxiliary problem now corresponds to a sphere steadily translating with unit velocity in the $\textbf{\textit{e}}_{\bf{1}}$-direction. Solving this problem with the techniques described in Appendix \ref{CompleA} yields an approximation of the corresponding velocity field, $\hat{\textbf{\textit{U}}}_{\parallel}$, accurate up to terms of $\mathcal{O}(\kappa^3)$. The surface quantities $\hat{\textbf{\textit{F}}}_{D\parallel}$, $\hat{\textbf{\textit{T}}}_{D\parallel}$ and $\hat{\textbf{\textit{S}}}_{D\parallel}$  which are the counterparts of $\hat{\textbf{\textit{F}}}_{D}$, $\hat{\textbf{\textit{T}}}_{D}$ and $\hat{\textbf{\textit{S}}}_{D}$ in (\ref{recipr}) may then be deduced; the corresponding evaluations result in (\ref{FDp})-(\ref{mom2p}). Last, the radial component of the inertial body force $\int_\mathcal{V_A}\frac{D\textbf{\textit{U}}_0}{Dt}d\mathcal{V}$ due to the undisturbed flow acceleration is
\begin{equation}
\textbf{\textit{F}}_0\cdot{\textbf{\textit{e}}_{\bf{1}}}=\frac{4}{3}\pi Re\left\{ \left(\alpha\frac{d\textbf{\textit{V}}_{S0}}{dt}+ \alpha_b\textbf{\textit{V}}_{S0}\right)\cdot{\textbf{\textit{e}}_{\bf{1}}}+\frac{1}{5}\left(\frac{d(\rho_0\alpha_d)}{dt}-3\rho_0\alpha_b\alpha_d\right)\right\}\,.
\end{equation}
\subsection{Stokes-flow approximation}
\label{SFAp}
Applying (\ref{mom1p}) and (\ref{mom2p}) to (\ref{carflow5p}), the $\textbf{\textit{e}}_{\bf{1}}$-projection of the reciprocal theorem indicates that the $\rho_0$-dependent radial component of the carrying flow generates a nonzero force such that
\begin{equation}
\textbf{\textit{F}}_{F\delta}\cdot\textbf{\textit{e}}_{\bf{1}}\approx-\frac{3}{8}\pi\alpha_c\rho_0\kappa\left\{5\kappa+\frac{8}{1+\Lambda}(1+\frac{9}{16}\kappa)\right\}\,.
\label{FSp}
\end{equation}
Both terms in the right-hand side of (\ref{FSp}) provide a negative contribution to $\textbf{\textit{F}}_{F\delta}$, making the particle lag behind the fluid in the $\textbf{\textit{e}}_{\bf{1}}$-direction.  Balancing (\ref{FSp}) with the drag force $-\hat{\textbf{\textit{F}}}_{D\parallel}\cdot\textbf{\textit{V}}_{S0}$ evaluated with the aid of (\ref{FDp}), the creeping-flow approximation indicates that, for small $\kappa$, the radial slip velocity is primarily due to the first term in the right-hand side of (\ref{FSp}). This yields
\begin{equation}
\textbf{\textit{V}}_{S0}\cdot\textbf{\textit{e}}_{\bf{1}}\approx-\alpha\rho_0\kappa^2\frac{\Lambda^2}{(1+\Lambda)^4}\,.
\label{slipp}
\end{equation}
Since $\Lambda=\kappa\Delta/k_\delta$, 
the radial slip in (\ref{slipp}), which originates from the curvature-induced Faxén force, is of $\mathcal{O}(\kappa^4\Delta^2)$ compared to the radial component of the primary straining flow. As the particle gets closer to the wall, $\Lambda$ becomes large. There, the dominant contribution to the right-hand side of (\ref{FSp}) is provided by the second term, \textit{i.e.} the wall-induced Faxén force associated with the radial shear flow $\rho_0\alpha_cx_3\textbf{\textit{e}}_{\bf{1}}$ in (\ref{carflow5p}), and the relative slip becomes of $\mathcal{O}(\kappa^2\Delta^{-1})$.
\subsection{Inertial corrections}
\label{intertpar}
Similar to the route followed in \S\,\ref{inerteff}, we first compute inertial forces due to unsteadiness and then consider advective contributions. 

First of all, the radial component of the force $\textbf{\textit{F}}_U$ due to possible time variations in the radial slip velocity was computed in M1 and was found to be
\begin{equation}
{\textbf{\textit{F}}}_U\cdot\textbf{\textit{e}}_{\bf{1}}=-\frac{9}{4}\pi \alpha Re\left(3\kappa^{-1}+\frac{217}{216}+\mathcal{O}(\kappa)\right)\frac{d\textbf{\textit{V}}_{S0}}{dt}\cdot\textbf{\textit{e}}_{\bf{1}}\,.
\end{equation}
 This result still applies here, as it is independent of the background flow. \vspace{2mm}\\
 \indent The argument provided in \S\,\ref{prelim} indicates that none of the inertial contributions resulting from the axisymmetric component of the carrying flow in (\ref{carflow5p}) can have a nonzero radial component. Hence, only the radial flow $\rho_0(t)\{\alpha_c(t)x_3+\alpha_d(t)x_3^2\}\textbf{\textit{e}}_{\bf{1}}$ in (\ref{carflow5p}) may provide nonzero radial forces arising from unsteadiness or advective transport. Moreover, contributions due to the parabolic component $\rho_0\alpha_dx_3^2\textbf{\textit{e}}_{\bf{1}}$ are smaller by a factor of $\mathcal{O}(\kappa(1+\Lambda)^{-1})$ than those due to the shear component $\rho_0\alpha_cx_3\textbf{\textit{e}}_{\bf{1}}$. Consequently, following the argument discussed in  \S\,\ref{unste}, only the latter needs to be considered at the present order of approximation. 
To compute the corresponding inertial corrections, the relevant shear Reynolds number has to be small. As the strength of the shear in (\ref{carflow5p}) is $\rho_0\alpha_c$ and the magnitude of $\alpha_c$ cannot exceed values of $\mathcal{O}(\alpha)$, this condition implies $\rho_0\alpha Re\ll1$, \textit{i.e.}
\begin{equation}
\rho_0\ll\Delta^2\,.
\end{equation}
Due to the presence of the radial shear component in the carrying flow, the disturbance now comprises a stresslet and an irrotational quadrupole which are not present when the particle stands on the axis of the HH flow. Close to the particle, the velocity disturbance induced by this stresslet, say $\textbf{\textit{u}}_{str\parallel}$, has the form $\frac{x_1x_3\textbf{\textit{x}}}{r^{5}}$ while that induced by the stresslet associated with the primary axisymmetric strain, say $\textbf{\textit{u}}_{str\perp}$, has the form $\frac{\textbf{\textit{x}}}{r^{3}}-3\frac{x_3^2\textbf{\textit{x}}}{r^{5}}$. 
\vspace{2mm}
\\
\indent Similar to (\ref{FI22}), the evolution of the radial and wall-normal particle positions result in a net force, as it makes the strength of the $\textbf{\textit{u}}_{str\parallel}$-contribution vary over time through the time variations of $\rho_0\alpha_c$. 
 Following the results and approximations discussed at the end of Appendix \ref{CIE}, the leading-order contribution to this force is found to be 
 \begin{equation}
{\textbf{\textit{F}}}_{U\delta}\cdot\textbf{\textit{e}}_{\bf{1}}\approx\frac{33}{4}\pi\alpha\rho_0\kappa^3\frac{7+2\Lambda}{k_\delta^2(1+\Lambda)^5}\,.
 \label{FUp}
 \end{equation}
Time variations of $\rho_0\alpha_d$ induce a qualitatively similar contribution, but it is negligible at the present order of approximation for the reason mentioned above.\vspace{2mm}\\
 \indent Let us now consider advective contributions. Gradients of the axisymmetric disturbance $\textbf{\textit{u}}_{str\perp}$ are advected by the shear flow and \textit{vice versa}, which yields a radial inertial force, say ${\textbf{\textit{F}}}_{I\delta}\cdot\textbf{\textit{e}}_{\bf{1}}$. As reported in Appendix \ref{CIE}, evaluation of (\ref{FI0p}) yields  
  \begin{equation}
 {\textbf{\textit{F}}}_{I\delta}\cdot\textbf{\textit{e}}_{\bf{1}}=\frac{15}{16}\pi\alpha\frac{\rho_0}{k_\delta^2}\kappa^3(1+\frac{9}{16}\kappa)\frac{(1+2\Lambda)}{(1+\Lambda)^5}\,.
 \label{FIp}
 \end{equation}
Here also we disregard the $\mathcal{O}(\kappa/(1+\Lambda))$-smaller contribution of the parabolic radial flow component in (\ref{carflow5p}) to the advective transport of $\textbf{\textit{u}}_{str\perp}$.\\
 \indent Similar to (\ref{FDA}) in the wall-normal direction, advection of the Stokeslet-type disturbance associated with the radial slip velocity  $\textbf{\textit{V}}_{S0}\cdot\textbf{\textit{e}}_{\bf{1}}$ by the base straining flow (and \textit{vice versa}) results in an inertial correction to the radial drag coefficient. According to (\ref{FDA0p}) and the comments that follow, evaluation of this contribution up to $\mathcal{O}(\kappa^0)$-terms yields
\begin{equation}
{\textbf{\textit{F}}}_{D\alpha}\cdot\textbf{\textit{e}}_{\bf{1}}=-\frac{\pi}{32}\alpha Re\frac{(1+2\Lambda)}{(1+\Lambda)^2}\left(99\kappa^{-1}+\frac{29237}{120}+\mathcal{O}(\kappa)\right)\textbf{\textit{V}}_{S0}\cdot\textbf{\textit{e}}_{\bf{1}}\,.
\label{FDAp}
\end{equation}
 Similarly, we must consider the force resulting from the transport of the same disturbance by the radial shear flow and \textit{vice versa}. However, the eigenvectors of the velocity gradient  $\textbf{\textit{e}}_{\bf{3}}\textbf{\textit{e}}_{\bf{1}}$ corresponding to the radial shear flow are inclined by an angle of $\pm\pi/4$ with respect to the $(\textbf{\textit{e}}_{\bf{1}},\,\textbf{\textit{e}}_{\bf{3}})$ axes. For this reason, this advective transport results in a transverse force along the $\textbf{\textit{e}}_{\bf{3}}$-direction, not in a correction to the drag. For the same reason,  the transport of the disturbance associated with the wall-normal slip $\textbf{\textit{V}}_{S0}\cdot\textbf{\textit{e}}_{\bf{3}}$ by the shear flow yields a radial force along the $\textbf{\textit{e}}_{\bf{1}}$-direction. The first of these contributions was computed to leading order by \cite{Cox1977}, and to second order by Lovalenti in an appendix to \cite{Cherukat1994}. The second was computed in M1 and M2; its second-order term was amended by \cite{Magnaudet2004}. Making use of these results and noting that the shear strength in (\ref{carflow5p}) is $\rho_0\alpha_c$, the lift force resulting from both contributions may be written in the form
 \begin{equation}
  {\textbf{\textit{F}}}_{L\delta}=-\frac{9}{16}\pi\frac{\rho_0}{k_\delta^2}\kappa^2\frac{1}{(1+\Lambda)^3}\left\{\left(5+\frac{253}{432}\kappa\right)(\textbf{\textit{V}}_{S0}\cdot\textbf{\textit{e}}_{\bf{3}})\textbf{\textit{e}}_{\bf{1}}+\left(\frac{11}{3}+\frac{443}{144}\kappa\right)(\textbf{\textit{V}}_{S0}\cdot\textbf{\textit{e}}_{\bf{1}})\textbf{\textit{e}}_{\bf{3}}\right\}\,.
 \label{FLp}
 \end{equation}
\indent Last, in a shear flow, the stresslet $\textbf{\textit{u}}_{str\parallel}$ is known to induce an inertial force perpendicular to the streamlines, \textit{i.e.} a lift force acting in the $\textbf{\textit{e}}_{\bf{3}}$-direction. With a shear rate $\alpha$ and a particle free to rotate as it is here, this contribution, first computed at leading order by \cite{Cox1977}, yields a force $\frac{55}{96}\pi\alpha^2Re\textbf{\textit{e}}_{\bf{3}}+\mathcal{O}(\kappa)$. Considering again that the shear rate in (\ref{carflow5p}) is $\rho_0\alpha_c$ and taking into account  the $1+\frac{9}{8}\kappa$ multiplicative factor resulting from the reflection of the Stokeslet at stake, this lift force, say $\textbf{\textit{F}}_{L\alpha^2}$, is here
\begin{equation}
\textbf{\textit{F}}_{L\alpha^2}\cdot\textbf{\textit{e}}_{\bf{3}}\approx\frac{55}{24}\pi\alpha \left(\frac{\rho_0}{k_\delta}\right)^2\kappa^4(1+\frac{9}{8}\kappa)\frac{\Lambda^2}{(1+\Lambda)^6}\,.
\label{CS}
\end{equation}
Although (\ref{CS}) reveals a $\kappa^4$-dependence of $\textbf{\textit{F}}_{L\alpha^2}$, $\rho_0$ may become large, which makes this force potentially significant when $\kappa$ increases, as discussed below. 
\subsection{Final force balance}
\label{FFBP}
 \indent The contributions derived in \S\,\ref{SFAp} and \ref{intertpar} may finally be gathered to obtain the differential equation governing the evolution of the radial slip. Defining  $E_\Lambda=\frac{1}{1+\Lambda}$ and $F_\Lambda=\frac{7+2\Lambda}{(1+\Lambda)^{2}}$ and applying the same truncation rules as in \S\,\ref{inerteff}, this force balance may be recast in the form
 \begin{eqnarray}
 \label{Refini2pp}
&&9 \alpha Re\left(3\kappa^{-1}+\frac{115}{72}\right)\frac{d\textbf{\textit{V}}_{S0}}{dt}\cdot\textbf{\textit{e}}_{\bf{1}}\\
\nonumber
&&+24\left\{1+\frac{9}{16}\kappa+\frac{81}{256}\kappa^2+\frac{217}{4096}\kappa^3+\frac{33}{64}\alpha Re\left(\kappa^{-1}+\frac{34357}{11880}\right)A_\Lambda\right\}\textbf{\textit{V}}_{S0}\cdot\textbf{\textit{e}}_{\bf{1}}\\
\nonumber
&&\approx3\alpha\rho_0\kappa^2\left\{\kappa B_\Lambda\left(11F_\Lambda+\frac{5}{4}A_\Lambda\right)-D_\Lambda\left(5\kappa+8E_\Lambda(1+\frac{9}{16}\kappa)\right)\right\}-\frac{45}{4}\rho_0\kappa^2B_\Lambda\textbf{\textit{V}}_{S0}\cdot\textbf{\textit{e}}_{\bf{3}}\,,
\end{eqnarray}
with $A_\Lambda$, $B_\Lambda$ and $D_\Lambda$ as defined in (\ref{definl}). \\
Moreover, (\ref{CS}) and the $\textbf{\textit{e}}_{\bf{3}}$-projection of (\ref{FLp}) represent lift contributions which alter the evolution of the wall-normal slip velocity. More specifically, at an arbitrary radial position $\rho_0(t)$, the right-hand side of (\ref{Refini2}) is supplemented by the $\rho_0$-dependent inertial contribution  
 \begin{equation}
F_{L3\rho_0}=\rho_0\kappa^2B_\Lambda\left\{\frac{55}{6}\alpha \rho_0\kappa^2D_\Lambda(1+\frac{9}{8}\kappa) -\frac{9}{4}\left(\frac{11}{3}+\frac{443}{144}\kappa\right)\textbf{\textit{V}}_{S0}\cdot\textbf{\textit{e}}_{\bf{1}}\right\}\,.
 \label{FL3p}
 \end{equation}
 Terms involving the slip velocity in the right-hand side of (\ref{Refini2pp}) and (\ref{FL3p}) couple the evolution of the slip along the $\textbf{\textit{e}}_{\bf{1}}$- and $\textbf{\textit{e}}_{\bf{3}}$-axes. In a given direction, they tend to produce a slip with opposite sign in the perpendicular direction. This is similar to the familiar Saffman lift force \citep{Saffman1965} which drives a particle leading the fluid toward the low-velocity side in a shear flow. Unlike the situation noticed in (\ref{Refini2}), the inertial correction to the drag coefficient is positive in (\ref{Refini2pp}), similar to the usual Oseen correction. Inertial effects proportional to $\alpha\rho_0$ in (\ref{Refini2pp}) and (\ref{FL3p}) provide positive source terms that tend to make the particle lead the fluid. 
 However, present expressions for the inertial corrections are valid only for separations such that $\kappa\gtrsim\Delta^{-1}$, so that $\Lambda$ is of $\mathcal{O}(1)$ or larger. Because of this, negative (\textit{i.e.} inward) zero-Reynolds-number effects corresponding to the two types of Faxén forces already present in (\ref{FSp}) always dominate in the right-hand side of (\ref{Refini2pp}), and inertial forces (\ref{FUp}) and (\ref{FDAp}) are only able to reduce the relative inward motion between the particle and the fluid.\\
  In contrast, the first term in the right-hand side of (\ref{FL3p}), which results from the lift force (\ref{CS}), may become large when the radial distance increases, owing to its $\rho_0^2$-dependence. Since it behaves as $(\rho_0/\Delta^2)^2$ very close to the wall ($\Lambda\gg1$), it is of $\mathcal{O}(\Delta^{-1})$ for $\rho_0\sim\Delta^{3/2}$, similar to the two Faxén contributions that dominate the right-hand side of the wall-normal force balance (\ref{Refini2}). It even becomes the dominant source term if $\rho_0$ stands in the range $\Delta^{3/2}\ll\rho_0\ll\Delta^2$. Indeed, at such large radial distances, the shear flow component in (\ref{carflow5p}) has become larger than the base straining flow. For this reason, the particle motion in the $\textbf{\textit{e}}_{\bf{3}}$-direction is dominated by lift effects associated with the shear, rather than by the interaction of the axisymmetric straining flow with the wall. In other terms, what (\ref{Refini2}) supplemented with (\ref{FL3p}) describes is the wall-normal dynamics of a particle in a carrying flow which gradually evolves from a bi-axial straining flow at small $\rho_0$ to a nearly wall-parallel uniform shear flow at large $\rho_0$. While this wall-normal dynamics is initially primarily governed by the wall-induced and curvature-induced Faxén forces (\ref{faxen}) and (\ref{Faclaa}), it becomes eventually dominated by the inertial shear-induced lift force (\ref{CS}).
 \section{Concluding remarks}
 \label{conclu}
In this investigation, we made use of a suitable form of the reciprocal theorem to establish the force balance on a neutrally-buoyant spherical particle moving close to a flat wall in an axisymmetric stagnation-point flow. An algebraic representation of the carrying flow within the boundary layer allowed us to obtain an approximate representation of the undisturbed velocity field valid throughout the flow domain. The corresponding representation specifies how the background linearly varying straining motion gradually transitions to a quadratic wall-parallel flow. To apply an asymptotic approach, we considered particles with sizes much smaller than the boundary layer thickness and small-but-finite Reynolds numbers. We employed a reflection technique truncated after three reflections, which keeps the technical difficulty reasonable but restricts predictions to moderate wall-particle separations, in principle not smaller than the particle radius. Conversely, we focused on separations smaller than the boundary layer thickness to obtain the leading-order expression of inertial effects through a regular expansion procedure. \\
\indent When the particle stands on the flow axis, it is submitted to two antagonistic Faxén forces, one specific to near-wall linearly varying flows, the other generic to quadratic carrying flows. Nevertheless the former is always dominant when the separation decreases, which tends to make the particle lag the fluid. Inertial effects reinforce this tendency in two ways. On the one hand, the wall induces an asymmetry in the advective transport of the disturbance, which results in repelling inertial forces depending only on the local strain rate of the carrying flow and relative size of the particle with respect to the separation. On the other hand, inertial corrections tend to reduce the drag coefficient, thus enhancing the slip velocity with respect to the creeping-flow limit. Overall, the wall-normal slip increases sharply as the particle gets closer to the wall; 
 the larger the particle, the larger the slip velocity. Present predictions are quantitatively confirmed by comparisons with data resulting from fully-resolved simulations within the range of separations and particle sizes where asymptotic expressions for the various forces are expected to be relevant.\\
\indent When the particle is released some distance from the flow axis and stands within the boundary layer,  a radial component of the slip velocity develops. The two types of Faxén forces contribute to generate an inward radial slip which makes the particle lag the fluid. In contrast, inertial effects increase the drag coefficient and tend to make the particle lead the fluid. For this reason, the overall radial slip is lowered by finite-$Re$ corrections. In addition, the fluid velocity in the vicinity of the particle comprises a radial shear component, the magnitude of which increases linearly with the radial distance to the flow axis. The near-wall advective transport associated with this shear generates several distinct lift forces acting along both the radial and wall-normal directions. All of these lift contributions tend to enhance the corresponding slip velocity component. 
The strength of the radial shear grows at the expense of the wall-normal straining component of the carrying flow when the radial distance to the axis of the HH flow increases. Hence the particle surroundings transition gradually toward the more familiar wall-parallel shear flow configuration in which a neutrally-buoyant particle has long been known to lag the fluid and experience a repelling lift force.\vspace{2mm}\\
\indent It is obviously desirable to extend present results toward smaller and larger separations. Predictions taking into account inertial corrections were found to agree well with results of fully-resolved simulations down to gaps corresponding approximately to one third of the particle radius. Extension toward smaller gaps is required to incorporate lubrication effects and predict the late stages of the particle approach to the wall. Nevertheless, the reflection technique is unsuitable for such an extension, as the flow within a narrow gap can barely be viewed as a small or even moderate distortion of the base disturbance in an unbounded flow. An appropriate representation, such as the bipolar co-ordinates system, is known to allow the exact viscous solution to be computed down to a vanishingly small gap \citep{Brenner1961,Maude1961,Rallabandi2017}. Employing this representation to express nonlinear inertial effects is probably a viable approach to obtain predictions at low-but-finite Reynolds number down to the wall \citep{Cherukat1994}. In the opposite limit, determining how the various near-wall inertial forces vary with increasing separation is required to obtain a uniformly valid description of the rheology of a suspension of neutrally-buoyant particles in the prototypical configuration of the HH flow.This is especially necessary regarding the unsteady Oseen force, whose asymptotic expression exhibits an unphysical growth and eventually a divergence at large separations, an undesired behaviour which was found to limit the range of applicability of present predictions. To this aim, it is necessary to consider situations in which the wall stands in the outer region of the disturbance, which immediately introduces a singular perturbation problem. Use of matched asymptotic expansions in the spirit of the study by \cite{Vasseur1977} on the near-wall migration of a particle in a stagnant fluid should provide the way to deal with this transitional regime.
\appendix
\section{Derivation of the force balance (\ref{recipr})}
\label{appA}
The reciprocal theorem providing the force balance on a buoyant drop with an arbitrary viscosity moving in an arbitrary direction with respect to a planar wall in a linear flow was obtained in M1 (equation (8)). Although the extension to a quadratic flow and the specialization to the case of a rigid particle are straightforward, we provide the complete derivation in this appendix for the sake of self-consistency.

First, using the scalings and definitions introduced in \S\,\ref{scaling}, the undisturbed flow obeys
\begin{eqnarray}
&&\nabla\cdot\textbf{\textit{U}}_0=0\,;\,\,\nabla\cdot\boldsymbol{\Sigma}_0=Re\frac{D\textbf{\textit{U}}_0}{Dt}
\equiv Re\left\{\alpha\frac{\partial\textbf{\textit{U}}_0}{\partial t}+(\textbf{\textit{U}}_0-\textbf{\textit{V}})\cdot\nabla\textbf{\textit{U}}_0\right\}\,\,\mbox{in}\,\,\mathcal{V}\,,\\
\label{recipr0002}
&&\textbf{\textit{U}}_0=\textbf{0}\quad\mbox{on} \quad\mathcal{A}_w\,,
\label{recipr0000}
\end{eqnarray}
where $\boldsymbol{\Sigma}_0$ is the undisturbed stress tensor, $\mathcal{A}_w$ denotes the planar wall bounding the fluid domain $\mathcal{V}$, and the Lagrangian acceleration $D\textbf{\textit{U}}_0/Dt$ is expressed in the reference frame $(\mathcal{R})$ translating with the particle.

Let now $\textbf{\textit{U}}=\textbf{\textit{U}}_0+\textbf{\textit{u}}-\textbf{\textit{V}}$ be the relative fluid velocity with respect to the particle, $\textbf{\textit{u}}$ denoting the velocity disturbance and $\textbf{\textit{V}}$ the absolute translational velocity of the particle. In $(\mathcal{R})$, the `direct' problem governing $\textbf{\textit{U}}$ and the associated stress tensor $\boldsymbol\Sigma$ is 
\begin{eqnarray}
\label{recipr00}
&&\nabla\cdot\textbf{\textit{U}}=0\,;\quad\nabla\cdot\boldsymbol{\Sigma}=Re\left\{\alpha\frac{\partial\textbf{\textit{U}}}{\partial t}+\textbf{\textit{U}}\cdot\nabla\textbf{\textit{U}}\right\}\quad\mbox{in} \quad\mathcal{V}\,,\\
\label{recipr01}
&&\textbf{\textit{U}}=\textbf{0}\quad\mbox{on} \quad\mathcal{A}\,,\\
\label{recipr02}
&&\textbf{\textit{U}}+\textbf{\textit{V}}=\textbf{0}\quad\mbox{on} \quad\mathcal{A}_w\,;\quad\textbf{\textit{U}}+\textbf{\textit{V}}\rightarrow\textbf{\textit{U}}_0\quad\mbox{for} \quad||\textbf{\textit{x}}||\rightarrow\infty \,,
\label{recipr0}
\end{eqnarray}
where $\mathcal{A}$ denotes the particle surface, and $\textbf{\textit{x}}$ is the local distance to the particle centre. Equation (\ref{recipr01}) and the first of (\ref{recipr02}) express the no-slip condition on the particle (assuming that it does not rotate) and wall surfaces, respectively, while the second of (\ref{recipr02}) expresses the vanishing of the disturbance in the far field. Since $(\mathcal{R})$ is non-inertial, the pressure field involved in $\boldsymbol\Sigma$ includes a contribution $\alpha Re\,\textbf{\textit{x}}\cdot d\textbf{\textit{V}}/dt$ due to the complementary acceleration. \\
In the `auxiliary' problem, the particle is assumed to steadily translate with unit velocity $\textbf{\textit{e}}_{\bf{3}}$. The corresponding relative velocity $\hat{\textbf{\textit{U}}}$ and associated stress tensor $\hat{\boldsymbol\Sigma}$ obey
\begin{eqnarray}
&&\nabla\cdot\hat{\textbf{\textit{U}}}=0\,;\quad\nabla\cdot\hat{\boldsymbol{\Sigma}}=\textbf{0}\quad\mbox{in} \quad\mathcal{V}\,,\\
\label{recipr001}
&&\hat{\textbf{\textit{U}}}=\textbf{0}\quad\mbox{on} \quad\mathcal{A}\,,\\
\label{recipr002}
&&\hat{\textbf{\textit{U}}}+\textbf{\textit{e}}_{\bf{3}}=\textbf{0}\quad\mbox{on} \quad\mathcal{A}_w\,;\quad\hat{\textbf{\textit{U}}}+\textbf{\textit{e}}_{\bf{3}}\rightarrow\textbf{0}\quad\mbox{for} \quad||\textbf{\textit{x}}||\rightarrow\infty\,,
\label{recipr000}
\end{eqnarray}
In the direct problem, the particle is assumed to be neutrally buoyant, so that it experiences no net force. In contrast, it experiences a net drag $\hat{\textbf{\textit{F}}}_D$ in the auxiliary problem. Hence
\begin{equation}
\quad\int_\mathcal{A}\boldsymbol\Sigma\cdot\textbf{\textit{n}}d\mathcal{S}=\textbf{0}\,;\quad \hat{\textbf{\textit{F}}}_D=\int_\mathcal{A}\hat{\boldsymbol\Sigma}\cdot\textbf{\textit{n}}d\mathcal{S}\,,
\label{force0}
\end{equation}
with $\textbf{\textit{n}}$ is the unit normal to $\mathcal{A}$ directed into the fluid. \\
Introducing the surface $\mathcal{A}_\infty$ bounding the fluid domain at large distances from the particle and the outward unit normal $\textbf{\textit{n}}_e$ to $\mathcal{V}$ (with $\textbf{\textit{n}}_e=-\textbf{\textit{n}}$ on $\mathcal{A}$), one can form the surface integral $\bigintsss_{\mathcal{A}\cup\mathcal{A}_w\cup\mathcal{A_\infty}}{\left\{(\hat{\textbf{\textit{U}}}+\textbf{\textit{e}}_{\bf{3}})\cdot\boldsymbol\Sigma-(\textbf{\textit{U}}+\textbf{\textit{V}})\cdot\hat{\boldsymbol{\Sigma}}\right\}\cdot{\textbf{\textit{n}}_e}}d\mathcal{S}$. Transforming this integral with the aid of the divergence theorem then yields
\begin{eqnarray}
\nonumber
&&\hat{\textbf{\textit{F}}}_D\cdot\textbf{\textit{V}}+\int_{\mathcal{A}_w\cup\mathcal{A_\infty}}\left\{(\hat{\textbf{\textit{U}}}+\textbf{\textit{e}}_{\bf{3}})\cdot\boldsymbol\Sigma-(\textbf{\textit{U}}+\textbf{\textit{V}})\cdot\hat{\boldsymbol{\Sigma}}\right\}\cdot{\textbf{\textit{n}}_e}d\mathcal{S}\\
&=&Re\int_\mathcal{V}(\hat{\textbf{\textit{U}}}+\textbf{\textit{e}}_{\bf{3}})\cdot\left(\alpha\frac{\partial\textbf{\textit{U}}}{\partial t}+\textbf{\textit{U}}\cdot\nabla\textbf{\textit{U}}\right)d\mathcal{V}\,.
\label{recipr1}
\end{eqnarray}
Note that although (\ref{recipr01}) includes an additional term if the particle rotates, (\ref{recipr1}) is left unchanged by this rotation because the particle is only translating in the `auxiliary' problem, so that the corresponding torque is zero. \\
Noting that $\textbf{\textit{U}}+\textbf{\textit{V}}\rightarrow\textbf{\textit{U}}_0$ and $\boldsymbol{\Sigma}\rightarrow\boldsymbol{\Sigma}_0-\alpha Re\left(\textbf{\textit{x}}\cdot d\textbf{\textit{V}}/dt\right)\textbf{\textit{I}}$ for $||\textbf{\textit{x}}||\rightarrow\infty$ (with $\textbf{\textit{I}}$ the Kronecker delta), and making use of the no-slip condition on $\mathcal{A}_w$, the surface integral in (\ref{recipr1}) is seen to tend toward $\bigintsss_{\mathcal{A}_w\cup\mathcal{A_\infty}}\left\{(\hat{\textbf{\textit{U}}}+\textbf{\textit{e}}_{\bf{3}})\cdot\{\boldsymbol\Sigma_0-\alpha Re\left(\textbf{\textit{x}}\cdot d\textbf{\textit{V}}/dt\right)\textbf{\textit{I}}\}-\textbf{\textit{U}}_0\cdot\hat{\boldsymbol{\Sigma}}\right\}\cdot{\textbf{\textit{n}}_e}d\mathcal{S}$.
Further use of the divergence theorem and the no-slip condition on $\mathcal{A}$ allows this surface integral to be transformed as
\begin{eqnarray}
\nonumber
&&\int_{\mathcal{A}_w\cup\mathcal{A_\infty}}\left\{(\hat{\textbf{\textit{U}}}+\textbf{\textit{e}}_{\bf{3}})\{\cdot\boldsymbol\Sigma_0-\alpha Re\left(\textbf{\textit{x}}\cdot d\textbf{\textit{V}}/dt\right)\textbf{\textit{I}}\}-\textbf{\textit{U}}_0\cdot\hat{\boldsymbol{\Sigma}}\right\}\cdot{\textbf{\textit{n}}_e}d\mathcal{S}\\
\nonumber
&=&Re\int_\mathcal{V}(\hat{\textbf{\textit{U}}}+\textbf{\textit{e}}_{\bf{3}})\cdot\left(\frac{D\textbf{\textit{U}}_0}{Dt}-\alpha\frac{d\textbf{\textit{V}}}{dt}\right)d\mathcal{V}+\int_\mathcal{A}\left\{\textbf{\textit{e}}_{\bf{3}}\cdot\boldsymbol\Sigma_0-\textbf{\textit{U}}_0\cdot\hat{\boldsymbol{\Sigma}}\right\}\cdot{\textbf{\textit{n}}}d\mathcal{S}\\
&&-\frac{4}{3}\pi\alpha Re\,\textbf{\textit{e}}_{\bf{3}}\cdot\frac{d\textbf{\textit{V}}}{dt}\,.
\label{recipr22}
\end{eqnarray}
Last, from the definition of $\textbf{\textit{U}}$ it is readily established that (see also equation (5) in M1 and the comments that follow)
\begin{equation}
\alpha\frac{\partial\textbf{\textit{U}}}{\partial t}+\textbf{\textit{U}}\cdot\nabla\textbf{\textit{U}}=\alpha\frac{\partial\textbf{\textit{u}}}{\partial t}+\textbf{\textit{U}}\cdot\nabla\textbf{\textit{u}}+\textbf{\textit{u}}\cdot\nabla\textbf{\textit{U}}_0+\frac{D\textbf{\textit{U}}_0}{Dt}-\alpha\frac{d\textbf{\textit{V}}}{dt}\,.
\label{recipr33}
\end{equation}
Introducing (\ref{recipr22}) in (\ref{recipr1}) and making use of (\ref{recipr33}) one finally obtains
\begin{eqnarray}
\nonumber
\frac{4}{3}\pi\alpha Re\,\textbf{\textit{e}}_{\bf{3}}\cdot\frac{d\textbf{\textit{V}}}{dt}&=&Re\textbf{\textit{e}}_{\bf{3}}\cdot\int_\mathcal{V_A}\frac{D\textbf{\textit{U}}_0}{Dt}d\mathcal{V}+\hat{\textbf{\textit{F}}}_D\cdot\textbf{\textit{V}}-\int_\mathcal{A}\textbf{\textit{U}}_0\cdot\hat{\boldsymbol{\Sigma}}\cdot{\textbf{\textit{n}}}d\mathcal{S}\\
&-&Re\int_\mathcal{V}(\hat{\textbf{\textit{U}}}+\textbf{\textit{e}}_{\bf{3}})\cdot\left(\alpha\frac{\partial\textbf{\textit{u}}}{\partial t}+\textbf{\textit{U}}\cdot\nabla\textbf{\textit{u}}+\textbf{\textit{u}}\cdot\nabla\textbf{\textit{U}}_0\right)d\mathcal{V}\,,
\label{recipr11}
\end{eqnarray}
where $\int_\mathcal{V_A}d\mathcal{V}=\frac{4}{3}\pi$ is the particle volume, $\mathcal{V_A}$ denoting the volume enclosed in $\mathcal{A}$.\\
To compute the surface integral in (\ref{recipr11}), we introduce a Taylor expansion of the undisturbed velocity about the particle centre in the form 
\begin{equation}
\textbf{\textit{U}}_0(\textbf{\textit{x}},t)=\textbf{\textit{U}}_0^0(t)+(\textbf{\textit{x}}\cdot\nabla^0)\textbf{\textit{U}}_0(t)+\frac{1}{2}(\textbf{\textit{x}}\textbf{\textit{x}}:\nabla^0\nabla)\textbf{\textit{U}}_0(t)+...\,,
\label{expan}
\end{equation}
where $\nabla^0\textbf{\textit{U}}_0(t)$ and $\nabla^0\nabla\textbf{\textit{U}}_0(t)$ denote the gradient and Hessian of the undisturbed velocity evaluated at the centre of the particle, respectively. Then, defining the particle slip velocity $\textbf{\textit{V}}_{S0}=\textbf{\textit{V}}-\textbf{\textit{U}}_0^0$ and the first- and second-order surface moments of the auxiliary surface traction $\hat{\boldsymbol{\Sigma}}\cdot{\textbf{\textit{n}}}$ as
\begin{equation}
\hat{\textbf{\textit{T}}}_D=\int_\mathcal{A}\textbf{\textit{x}}\hat{\boldsymbol{\Sigma}}\cdot{\textbf{\textit{n}}}d\mathcal{S}\,;\quad
 \hat{\textbf{\textit{S}}}_D=\int_\mathcal{A}\textbf{\textit{x}}\textbf{\textit{x}}\hat{\boldsymbol{\Sigma}}\cdot{\textbf{\textit{n}}}d\mathcal{S}\,,
 \label{moments}
 \end{equation}
(\ref{recipr}) is obtained.\\

\section{Approximate solution of the auxiliary problem}
\label{CompleA}
An approximate solution of the auxiliary problem may be sought in the form of a series of `reflections' of the fundamental solution corresponding to a particle translating in an unbounded fluid. The solution is expanded with respect to the small parameter $\kappa$, the inverse of the dimensionless distance separating the particle from the wall. At $\mathcal{O}(\kappa^0)$, the fundamental solution satisfying the no-slip condition at the particle surface is the sum of a Stokeslet and an irrotational dipole (or degenerate Stokes quadrupole). These singularities induce velocity disturbances decaying with the distance $r=||\textbf{\textit{x}}||$ to the particle centre as $r^{-1}$ and $r^{-3}$, respectively. Therefore the remains of these disturbances are of $\mathcal{O}(\kappa)$ and $\mathcal{O}(\kappa^3)$ at the wall, respectively. To satisfy the no-slip condition there, image singularities have to be added to the solution. Determining these images is made possible by using Faxén's transformation which allows an integral representation of fundamental solutions of the Laplace equation in the presence of a wall \citep{Happel1973,Ho1974}. Image solutions can then be expanded in the vicinity of the particle to determine the wall-induced disturbance `felt' by the latter. Following this technique, the image of the fundamental Stokeslet is found to induce the near-particle disturbance $-\frac{9}{8}\kappa{\textbf{\textit{e}}_{\bf{3}}}-\frac{9}{32}\kappa^2({\textbf{\textit{x}}}-3x_3\textbf{\textit{e}}_{\bf{3}})+\mathcal{O}(\kappa^3)$. This disturbance implies that a Stokeslet with strength $\frac{27}{32}\kappa$ and a stresslet with strength $\frac{45}{64}\kappa^2$, plus associated irrotational dipoles and quadrupoles, have to be added to the fundamental solution to enforce the no-slip boundary condition at the particle surface. Successive reflections may be carried out to further improve the representation as the particle gets close to the wall. The drag force $\hat{\textbf{\textit{F}}}_D$ and the first- and second-order moments $\hat{\textbf{\textit{T}}}_D$ and $\hat{\textbf{\textit{S}}}_D$ involved in (\ref{recipr}) may finally be computed, which yields (see equations (A6) and (A7) in M1 for $\hat{\textbf{\textit{F}}}_D$ and $\hat{\textbf{\textit{T}}}_D$, respectively) 
\begin{eqnarray}
\label{comple1}
\hat{\textbf{\textit{F}}}_D&=&-6\pi(1+\frac{9}{8}\kappa+\frac{81}{64}\kappa^2+\frac{473}{512}\kappa^3+...)\textbf{\textit{e}}_{\bf{3}}+\mathcal{O}(\kappa^4)\,,\\
\label{comple2}
 \hat{\textbf{\textit{T}}}_D&=&-\frac{15}{8}\pi\kappa^2(1+\frac{9}{8}\kappa+...)({\textbf{\textit{e}}_1\textbf{\textit{e}}_1}+{\textbf{\textit{e}}_2\textbf{\textit{e}}_2}-2{\textbf{\textit{e}}_3\textbf{\textit{e}}_3})+\mathcal{O}(\kappa^4)\,,\\
 \label{comple3}
\hat{\textbf{\textit{S}}}_D&=&-2\pi(1+\frac{9}{8}\kappa+\frac{81}{64}\kappa^2+\frac{217}{512}\kappa^3){\textbf{\textit{I}}}\textbf{\textit{e}}_{\bf{3}}-\frac{15}{4}\pi\kappa^3\textbf{\textit{e}}_{\bf{3}}\textbf{\textit{e}}_{\bf{3}}\textbf{\textit{e}}_{\bf{3}}+\mathcal{O}(\kappa^4)\,.
\label{comple}
\end{eqnarray}
Note that the second-order moment $\hat{\textbf{\textit{S}}}_D=\int_\mathcal{A}\textbf{\textit{x}}\textbf{\textit{x}}(\hat{\boldsymbol{\Sigma}}\cdot{\textbf{\textit{e}}_r})d\mathcal{S}$ (a third-order tensor) remains isotropic on its first two indices only up to $\mathcal{O}(\kappa^2)$. 
  At next order, the $\mathcal{O}(\kappa^3)$-image of the fundamental Stokeslet induces a quadratic correction $\frac{3}{16}\kappa^3\{x_3{\textbf{\textit{x}}}+(\frac{5}{2}(x_1^2+x_2^2)-2x_3^2)\textbf{\textit{e}}_{\bf{3}}\}$ in the near-particle flow. This correction and the associated singularities (Stokes quadrupole, Stokeslet, irrotational octupole and dipole) yield the $-\frac{15}{4}\pi\kappa^3\textbf{\textit{e}}_{\bf{3}}\textbf{\textit{e}}_{\bf{3}}\textbf{\textit{e}}_{\bf{3}}$ contribution in (\ref{comple3}). \\
  In M2 it was pointed out that the $\mathcal{O}(\kappa^5)$-approximation of $\hat{\textbf{\textit{F}}}_D$ predicts an infinite drag for $\kappa\approx0.85$, while the exact solution \citep{Brenner1961,Maude1961} proves that the drag remains finite until the particle touches the wall, \textit{i.e.} $\kappa=1$. This is because in the unbounded solution which serves as a starting point for the reflection technique, streamlines exhibit a fore-aft symmetry past the particle, while for $\kappa\lesssim1$ the actual streamlines in the gap are highly distorted by the presence of the wall. This remark gives an indication regarding the minimum gap for which the reflection technique provides a satisfactory approximation of the near-wall disturbance. Based on a comparison with full numerical solutions, its was concluded in M2 that the $\mathcal{O}(\kappa^5)$-approximation allows a realistic estimate of $\hat{\textbf{\textit{F}}}_D$ up to $\kappa\approx0.7$. With $\kappa=0.5$, the $\mathcal{O}(\kappa^3)$-approximation in (\ref{comple1}) predicts that the drag is $1.995$ times larger than in an unbounded flow, while the aforementioned $\mathcal{O}(\kappa^5)$-approximation (equation  (51b) in M2) predicts an increase by a factor of 2.16, very close to the exact solution displayed in figure 3 of \cite{Rallabandi2017} which yields a factor of $2.14$. Hence the $\mathcal{O}(\kappa^3)$-prediction is within $7\%$ of the actual drag, and this difference decreases to less than $3\%$ for $\kappa=0.4$. These estimates indicate that the $\mathcal{O}(\kappa^4)$-truncation of the solution of the auxiliary problem provides accurate predictions for the drag force for $\kappa\lesssim0.5$.\vspace{2mm} \\
In \S\,\ref{Off}, the solution of the auxiliary problem corresponding to a particle steadily translating with unit velocity in the $\textbf{\textit{e}}_{\bf{1}}$-direction is involved. This solution, which we denote with a $_\parallel$ index, may be found in M1 (equations (A3$a$), (A5) and (A7$a$)) and M2 (equations (13b), (C2), (C3)). In particular one has
\begin{eqnarray}
\label{FDp}
\hat{\textbf{\textit{F}}}_{D\parallel}&=&\int_\mathcal{A}(\hat{\boldsymbol{\Sigma_\parallel}}\cdot{\textbf{\textit{n}}})d\mathcal{S}=-6\pi(1+\frac{9}{16}\kappa+\frac{81}{256}\kappa^2+\frac{217}{4096}\kappa^3)\textbf{\textit{e}}_{\bf{1}}+\mathcal{O}(\kappa^4)\,,\\
\label{mom1p}
 \hat{\textbf{\textit{T}}}_{D\parallel}&=&\int_\mathcal{A}\textbf{\textit{x}}(\hat{\boldsymbol{\Sigma_\parallel}}\cdot{\textbf{\textit{n}}})d\mathcal{S}=\frac{15}{8}\pi\kappa^2(1+\frac{9}{16}\kappa)({\textbf{\textit{e}}_1\textbf{\textit{e}}_3}+{\textbf{\textit{e}}_3\textbf{\textit{e}}_1})+\mathcal{O}(\kappa^4)\,,\\
 \label{mom2p}
 \hat{\textbf{\textit{S}}}_{D\parallel}&=&\int_\mathcal{A}\textbf{\textit{x}}\textbf{\textit{x}}(\hat{\boldsymbol{\Sigma_\parallel}}\cdot{\textbf{\textit{n}}})d\mathcal{S}=-2\pi(1+\frac{9}{16}\kappa+\frac{81}{512}\kappa^2){\textbf{\textit{I}}}\textbf{\textit{e}}_{\bf{1}}+\mathcal{O}(\kappa^3)\,,
\label{complep}
\end{eqnarray}
where the first- and second-order moments $\hat{\textbf{\textit{T}}}_{D\parallel}$ and $\hat{\textbf{\textit{S}}}_{D\parallel}$ of the surface traction $\hat{\boldsymbol{\Sigma_\parallel}}\cdot{\textbf{\textit{n}}}$ are required to evaluate the wall- and curvature-induced Fax\'en forces, respectively. 
\section{Technical characteristics of fully-resolved simulations}
\label{Simuli}
\indent The numerical results which serve as a reference to check the present predictions were obtained with fully-resolved simulations based on the axisymmetric time-dependent Navier-Stokes equations. Technical details are provided in \cite{Li2019} and only a brief summary is given here for the sake of self-consistency. \\
\indent The Navier-Stokes solver is based on a finite-volume spatial discretization on a staggered grid, with spatial derivatives evaluated using centered schemes. A third-order Runge-Kutta Crank-Nicolson time-advancement algorithm coupled with a projection technique is employed to advance the solution in time and satisfy the incompressibility condition.  An immersed boundary technique is used to determine the particle position as a function of time. To this end, an artificial force density is added to the fluid momentum equation. This force is set to zero outside the particle using a smoothed Heaviside function. Within the volume occupied by the particle, it is proportional to the difference between the local fluid velocity and the particle velocity, and inversely proportional to the time step. In this way, it enforces the no-slip boundary condition at the particle surface.   
The particle motion is governed by Newton's second law. The coupling between the flow solver and the immersed boundary scheme is achieved by expressing the hydrodynamic force on the particle as the difference between the time rate-of-change of the fluid momentum enclosed within the particle volume and the volume integral of the above artificial force. 
\\
\indent The simulations are carried out within a cylindrical domain with a size of $32 \delta\times63\delta$ (with $\delta=(\nu/B)^{1/2}$) in the radial and wall-normal directions, respectively. The velocity components corresponding to the theoretical Homann solution \citep{Homann1936} are imposed on all boundaries of this domain, except on the bounding wall where the fluid velocity is set to zero. Particles are released from rest on the flow axis at a position such that the initial dimensionless gap is $\epsilon_i=30$ in each case. Thus, the initial wall-particle separation ranges from $9.3\delta$ for the smallest particle to $15.5\delta$ for the largest one. In all cases, the particles quickly adjust to the carrying flow, so that their slip velocity is reduced to negligibly small values well before they enter the boundary layer.\\
The computational grid is highly nonuniform, being much refined in the wall-normal direction near the stagnation point to capture lubrication effects. For the three particle sizes considered in  \S\S\,\ref{compaDNS} and \ref{compan}, the minimum cell size is $1.5\times10^{-3}\delta$ in the radial direction close to the flow axis, and $1\times10^{-4}\delta$ in the wall-normal direction close to the wall. Over one particle radius, the number of grid cells in the radial direction ranges from $32$ for the smallest particle to $43$ for the largest one. In the wall-normal direction, this number depends on the particle position, increasing as the separation decreases. When the wall-particle gap equals the particle radius ($\epsilon=1$), it ranges from $33$ for the smallest particle to $46$ for the largest one.  It is important to stress that properly capturing the particle-induced disturbance in the present neutrally-buoyant situation requires a significantly finer grid than in the more familiar buoyancy/gravity-driven case. This is because, close to the particle, the disturbance decays as $1/r^2$ with the distance to the particle centre, instead of $1/r$ in the latter case. 
\section{Computation of near-wall inertial effects}
\label{CIE}
The procedure required to compute inertial corrections in the framework of the present assumptions was established  by \cite{Cox1968} (see section 6.1 in M2 for a summary).  First of all, it is convenient to introduce the outer co-ordinates $(\overline{x}_1,\overline{x}_2,\overline{x}_3)=\kappa(x_1,x_2,x_3)$, so that the wall stands at $\overline{x}_3=-1$ and the particle is shrunk to a small sphere $\overline{r}\leq\kappa$ around the origin $\overline{{\textbf{\textit{x}}}}={\bf{0}}$. With these strained co-ordinates, the elementary volume is  $d\overline{\mathcal{V}}=\kappa^{-3}d\mathcal{V}$ and the gradient operator is changed into $\overline{\nabla}=\kappa^{-1}\nabla$. Then a uniformly valid approximation of the leading contributions to the velocity fields $\hat{\textbf{\textit{U}}}$ and $\textbf{\textit{u}}$ involved in (\ref{recipr}) is required. This approximation, which we denote as $\overline{\hat{\textbf{\textit{U}}}}$ and $\overline{\textbf{\textit{u}}}$, respectively, has to satisfy the no-slip condition on both the particle and wall. \vspace{1mm}\\
\indent We detail the procedure in the case of the forces $ {\textbf{\textit{F}}}_{U\delta}$ and ${\textbf{\textit{F}}}_{I}$ encountered in \S\S\,\ref{unste} and \ref{adve}, respectively; the evaluation of all other inertial contributions follows a similar path. As the fundamental contribution to $\overline{\hat{\textbf{\textit{U}}}}$ (resp. $\overline{\textbf{\textit{u}}}$) is a Stokeslet (resp. stresslet) plus the corresponding image, they are respectively of $\mathcal{O}(\kappa)$ and $\mathcal{O}(\alpha_b\kappa^{2})$ once expressed in strained co-ordinates. The corresponding pre-factors are $\frac{3}{4}$ and $-\frac{5}{2}\alpha_b$, respectively (\textit{e.g.} equations (A2$a$) and (A2$c$) in M1). Therefore, referring to (\ref{recipr}), the leading-order contribution to $ {\textbf{\textit{F}}}_{U\delta}$, say $ {\textbf{\textit{F}}}_{U\delta0}$, may be written as
\begin{equation}
 {\textbf{\textit{F}}}_{U\delta0}\cdot\textbf{\textit{e}}_{\bf{3}}\approx\frac{15}{8}\alpha Re\frac{d\alpha_b}{dt}(\textbf{\textit{U}}_0^0+\textbf{\textit{V}}_{S0})\cdot\textbf{\textit{e}}_{\bf{3}}\int_{\overline{\mathcal{V}}}\overline{\hat{\textbf{\textit{U}}}}_{Sto}\cdot\overline{\textbf{\textit{U}}}_{str}d\overline{\mathcal{V}}\,,
 \label{FI200}
 \end{equation} 
  where $\overline{\hat{\textbf{\textit{U}}}}_{Sto}$ (resp. $\overline{\textbf{\textit{U}}}_{Str}$) stands for the uniformly valid expression of the unit Stokeslet (resp. stresslet) plus its image. 
 Similarly, based on (\ref{carflow5}) and (\ref{recipr}), the leading contribution ${\textbf{\textit{F}}}_{I0}$ to ${\textbf{\textit{F}}}_{I}$ is
 \begin{equation}
 {\textbf{\textit{F}}}_{I0}\cdot\textbf{\textit{e}}_{\bf{3}}=\frac{15}{8}\alpha_b^2 Re\int_{\overline{\mathcal{V}}}\overline{\hat{\textbf{\textit{U}}}}_{Sto}\cdot\left\{\overline{\textbf{\textit{U}}}_{str}\cdot({\textbf{\textit{I}}}-3\textbf{\textit{e}}_{\bf{3}}\textbf{\textit{e}}_{\bf{3}})+({\overline{{\textbf{\textit{x}}}}}-3\overline{x}_3\textbf{\textit{e}}_{\bf{3}})\cdot\overline{\nabla}\overline{\textbf{\textit{U}}}_{str}\right\}d\overline{\mathcal{V}}\,.
 \label{FI1}
 \end{equation}
 Following the techniques outlined in appendix C of M2, one finds
  \begin{eqnarray}
 \nonumber
 \overline{\hat{\textbf{\textit{U}}}}_{Sto}&=&\left(\frac{1}{\overline{r}}-\frac{1}{\tau}\right)\textbf{\textit{e}}_{\bf{3}}+\left(\frac{1}{\overline{r}^3}-\frac{1}{\tau^3}\right)\overline{x}_3\overline{{\textbf{\textit{x}}}}-2\frac{(1+\overline{x}_3)}{\tau^3}\left(\textbf{\textit{e}}_{\bf{3}}+3\frac{(2+\overline{x}_3)}{\tau^2}(\overline{{\textbf{\textit{x}}}}+2\textbf{\textit{e}}_{\bf{3}})\right)\,,\\
 \nonumber
  \overline{\textbf{\textit{U}}}_{str}&=&\left(\frac{1}{\overline{r}^3}-\frac{1}{\tau^3}\right)\overline{{\textbf{\textit{x}}}}-3\left(\frac{1}{\overline{r}^5}-\frac{1}{\tau^5}\right)\overline{x}_3^2\overline{{\textbf{\textit{x}}}}\\
 &+&6\frac{(1+\overline{x}_3)}{\tau^5}\left(2\overline{x}_3\textbf{\textit{e}}_{\bf{3}}+3(\overline{{\textbf{\textit{x}}}}+2\textbf{\textit{e}}_{\bf{3}})-5\frac{(2+\overline{x}_3)^2}{\tau^2}(\overline{{\textbf{\textit{x}}}}+2\textbf{\textit{e}}_{\bf{3}})\right)\,,
 \label{outer}
 \end{eqnarray}
 with $\overline{r}=(\overline{x}_1^2+\overline{x}_2^2+\overline{x}_3^2)^{1/2}$ and $\tau=(\overline{r}^2+4(1+\overline{x}_3))^{1/2}$ (note that $\tau=\overline{r}$ for $\overline{x}_3=-1$, \textit{i.e.} at the wall, and $\tau>\overline{r}$ everywhere in the fluid domain). As both fields exhibit axial symmetry with respect to the $\overline{x_3}$-direction, the volume integrals in (\ref{FI200}) and (\ref{FI1}) may be reduced to double integrals, say $2\pi\mathcal{I}$ with $\mathcal{I}=\int_{-1}^{\infty}\int_0^{\infty}J(\overline{\rho},\overline{x}_3)\overline{\rho} d\overline{\rho} d\overline{x}_3$, by setting $\overline{r}=(\overline{\rho}^2+\overline{x}_3^2)^{1/2}$ and integrating along the azimuthal direction. The double integrals may presumably be evaluated exactly by employing contour integration. To save time, we rather evaluated them numerically using the open software Maxima, after having circumvented the integrable singularity at $\overline{\textbf{\textit{x}}}={\bf{0}}$. In the case of (\ref{FI200}), this evaluation returned $\mathcal{I}=-0.9999$ with a 4-digit accuracy, from which we inferred that the exact value is $-1$. Similarly, with the same accuracy, we found $\mathcal{I}=1.24998$ in the case of (\ref{FI1}), from which we inferred that the exact value is $\mathcal{I}=5/4$. Therefore (\ref{FI200}) and (\ref{FI1}) yield eventually
 \begin{eqnarray}
  \label{FI20}
 {\textbf{\textit{F}}}_{U\delta0}\cdot\textbf{\textit{e}}_{\bf{3}}&\approx&-\frac{15}{4}\pi\alpha Re\frac{d\alpha_b}{dt}(\textbf{\textit{U}}_0^0+\textbf{\textit{V}}_{S0})\cdot\textbf{\textit{e}}_{\bf{3}}\,,\\
  \label{FI10}
 {\textbf{\textit{F}}}_{I0}\cdot\textbf{\textit{e}}_{\bf{3}}&\approx&\frac{75}{16}\pi\alpha_b^2 Re\,.
 \end{eqnarray}
 \noindent 
 Equations (\ref{FI20}) and (\ref{FI10}) only provide the leading-order term in the $\kappa$-expansion of the corresponding inertial force, say ${\textbf{\textit{F}}}_{U\delta}\cdot\textbf{\textit{e}}_{\bf{3}}$ and $ {\textbf{\textit{F}}}_{I}\cdot\textbf{\textit{e}}_{\bf{3}}$, respectively. In general, computing higher-order terms requires several additional contributions to be considered. First of all, the integration volume $\overline{\mathcal{V}}$ used during the numerical evaluation of (\ref{FI200}) and (\ref{FI1}) was artificially extended within the particle volume. Therefore the contribution provided by this volume must be subtracted from the result. Second, at $\mathcal{O}(\kappa^0)$ and $\mathcal{O}(\kappa)$, the complete velocity disturbance past the particle in the `direct' (resp. `auxiliary') problem involves a stresslet and an irrotational quadrupole (resp. a Stokeslet and an irrotational dipole). Contributions due to the two irrotational singularities are not accounted for in (\ref{FI200}) and (\ref{FI1}). They may be evaluated in unstrained co-ordinates by integrating the corresponding combinations of terms involved in the volume integrals $\int_\mathcal{V_I}(\hat{\textbf{\textit{U}}}+\textbf{\textit{e}}_{\bf{3}})\cdot(\partial\textbf{\textit{u}}/\partial t)d\mathcal{V}$ and 
 $\int_\mathcal{V_I}(\hat{\textbf{\textit{U}}}+\textbf{\textit{e}}_{\bf{3}})\cdot\left\{\textbf{\textit{u}}\cdot({\textbf{\textit{I}}}-3{\textbf{\textit{e}}_{\bf{3}}\textbf{\textit{e}}_{\bf{3}}})+({{\textbf{\textit{x}}}}-3x_3\textbf{\textit{e}}_{\bf{3}})\cdot\nabla\textbf{\textit{u}}\right\}d\mathcal{V}$, respectively. In these integrals, the relevant integration volume $\mathcal{V_I}$ is the `inner' fluid volume within which the distance to the particle centre is such that $1\le r<k_0\kappa^{-\gamma}$ with $k_0=\mathcal{O}(\kappa^0)$ and $0<\gamma<1$ \citep{Cox1968}. However, in the specific case of ${\textbf{\textit{F}}}_{U\delta}$ and $ {\textbf{\textit{F}}}_{I}$, both the disturbance $\textbf{\textit{u}}$ and (in the case of $ {\textbf{\textit{F}}}_{I}$) the straining component of the ambient velocity field are odd functions of $x_3$ close to the particle, up to $\mathcal{O}(\kappa^2)$-corrections. For this reason, all of the above terms result in a zero net contribution to the $\mathcal{O}(\kappa)$-correction of the corresponding force. In contrast, the magnitude of the Stokeslet $\overline{\hat{\textbf{\textit{U}}}}_{Sto}$ in (\ref{FI200}) and (\ref{FI1}) is actually $\frac{3}{4}(1+\frac{9}{8}\kappa+...)$, owing to its successive reflections. Consequently, the next term in the $\kappa$-expansion of these inertial forces is merely $\frac{9}{8}\kappa{\textbf{\textit{F}}}_{U\delta0}$ and $\frac{9}{8}\kappa{\textbf{\textit{F}}}_{I0}$, which finally yields (\ref{FI22}) and (\ref{FI}), respectively. 
  \vspace{2mm}\\
  \indent  Within the boundary layer, the advective transport of the $\mathcal{O}(\alpha_b)$-stresslet by the quadratic flow and \textit{vice versa} yields an additional $\mathcal{O}(\kappa^{-1}\alpha_b\alpha_cRe)$-force, which at leading order, is
\begin{eqnarray}
\nonumber
 {\textbf{\textit{F}}}_{I\delta0}\cdot\textbf{\textit{e}}_{\bf{3}}\approx
\frac{15}{8}\alpha_b\alpha_cRe\kappa^{-1} \int_{\overline{\mathcal{V}}}\overline{\hat{\textbf{\textit{U}}}}_{Sto}&\cdot&\bigg\{\overline{\textbf{\textit{U}}}_{str}\cdot\{\textbf{\textit{e}}_{\bf{3}}\overline{\textbf{\textit{x}}}+\overline{x}_3({\textbf{\textit{I}}}-4\textbf{\textit{e}}_{\bf{3}}\textbf{\textit{e}}_{\bf{3}})\}\bigg.\\
 &+&\bigg.\overline{x}_3({\overline{{\textbf{\textit{x}}}}}-2\overline{x}_3\textbf{\textit{e}}_{\bf{3}})\cdot\overline{\nabla}\overline{\textbf{\textit{U}}}_{str}\bigg\}d\overline{\mathcal{V}}\,.
 \label{FI3}
 \end{eqnarray}
 Numerical integration returned the value of the volume integral as $2\pi\times2.8333$, \textit{i.e.} virtually $\frac{17}{3}\pi$. Since all integrands involved in the first-order `inner' corrections to this leading-order estimate are even functions of $x_3$, they provide nonzero contributions at $\mathcal{O}(\alpha\kappa^3)$. Nevertheless, due to the definition of $\alpha_b$ and $\alpha_c$ in (\ref{coeff}), these contributions are weighted by $ \frac{1+2\Lambda}{(1+\Lambda)^5}$, whereas the $\mathcal{O}(\alpha\kappa^3)$-correction to the wall-induced Faxén force in (\ref{Rezerod}) is weighted by $ \frac{1+2\Lambda}{(1+\Lambda)^2}$. Following the argument discussed in \S\,\ref{unste}, the former corrections are negligibly small in the present context. Consequently, the  relevant approximation for the inertial force under consideration is merely ${\textbf{\textit{F}}}_{I\delta}\approx{\textbf{\textit{F}}}_{I\delta0}$, which yields (\ref{FI32}).\vspace{1mm}\\
\indent  The inertial correction to the drag coefficient arising from the transport of the Stokeslet associated with the slip velocity by the base straining flow and \textit{vice versa} may be computed though a similar approach. The formal expression for the leading term of this contribution, say ${\textbf{\textit{F}}}_{D\alpha0}$, is similar to that of ${\textbf{\textit{F}}}_{I0}$ in (\ref{FI1}), except that 
 $\overline{\textbf{\textit{U}}}_{str}$ has to be replaced by $\overline{\hat{\textbf{\textit{U}}}}_{Sto}$ and the pre-factor is now $-\frac{9}{16}\alpha_b Re\kappa^{-1}\textbf{\textit{V}}_{S0}\cdot\textbf{\textit{e}}_{\bf{3}}$.
 Using the technique outlined above, the volume integral was found to be $2\pi\times(-2.5001)$, from which we infer that its exact value is $-5\pi$, so that 
 \begin{equation}
 {\textbf{\textit{F}}}_{D\alpha0}\cdot\textbf{\textit{e}}_{\bf{3}}=\frac{45}{16}\pi\alpha_b Re\kappa^{-1}\textbf{\textit{V}}_{S0}\cdot\textbf{\textit{e}}_{\bf{3}}\,.
 \label{FDA0}
 \end{equation}
 In this case, the integrand is an even function of $x_3$ in the vicinity of the particle. Therefore the calculation of the $\mathcal{O}(\kappa^0)$-correction to ${\textbf{\textit{F}}}_{D\alpha0}$ requires the aforementioned `inner' terms to be evaluated. Moreover, the combination of the two Stokeslets at stake implies that the actual pre-factor of (\ref{FDA0}) is $\frac{45}{16}\pi\alpha_b Re\kappa^{-1}(1+\frac{9}{4}\kappa+...)$. Gathering all $\mathcal{O}(\kappa^0)$-terms eventually yields ${\textbf{\textit{F}}}_{D\alpha}\cdot\textbf{\textit{e}}_{\bf{3}}=(1+\frac{9}{4}\kappa){\textbf{\textit{F}}}_{D\alpha0}\cdot\textbf{\textit{e}}_{\bf{3}}-\frac{124}{15}\pi+\mathcal{O}(\kappa)$, which leads to (\ref{FDA}).\vspace{2mm}\\
 \indent  At leading order, the contribution resulting from the transport of the Stokeslet associated with the slip velocity by the quadratic flow and \textit{vice versa}, say ${\textbf{\textit{F}}}_{D\delta0}$, is similar to that of ${\textbf{\textit{F}}}_{I\delta0}$ in (\ref{FI3}), except that 
 $\overline{\textbf{\textit{U}}}_{str}$ has to be replaced by $\overline{\hat{\textbf{\textit{U}}}}_{Sto}$ and the pre-factor is now $-\frac{9}{8}\frac{\kappa}{k_\delta^2(1+\Lambda)^3}\textbf{\textit{V}}_{S0}\cdot\textbf{\textit{e}}_{\bf{3}}$. The value of the volume integral returned by numerical integration was $2\pi\times(-0.6666)\approx-\frac{4}{3}\pi$. All integrands involved in the first-order `inner' corrections are odd functions of $x_3$, so that the only contribution at $\mathcal{O}(\kappa^2)$ results from the reflection of the two Stokeslets, which yields a $(1+\frac{9}{4}\kappa)$-multiplicative factor. Neglecting $\mathcal{O}(\kappa^3)$-terms in agreement with the argument discussed in \S\,\ref{unste}, the $\mathcal{O}(\alpha\kappa^2)$-approximation of this force is then ${\textbf{\textit{F}}}_{D\delta}\approx(1+\frac{9}{4}\kappa){\textbf{\textit{F}}}_{D\delta0}$, from which (\ref{FDC}) is obtained.\vspace{2mm}\\
\indent  Inertial forces also affect the radial slip velocity when the particle stands some distance away from the flow axis. Their computation involves the uniformly valid expression of the unit Stokeslet in the $\textbf{\textit{e}}_{\bf{1}}$-direction and, for some of them, that of the unit stresslet associated with the shear component of the base radial flow. According to equations (C2) and (C5) in M2, the corresponding expressions are 
\begin{eqnarray}
\label{Stop}
 \overline{\hat{\textbf{\textit{U}}}}_{Sto\parallel}&=&\left(\frac{1}{\overline{r}}-\frac{1}{\tau}\right)\textbf{\textit{e}}_{\bf{1}}+\left(\frac{1}{\overline{r}^3}-\frac{1}{\tau^3}\right)\overline{x}_1\overline{{\textbf{\textit{x}}}}-2\frac{(1+\overline{x}_3)}{\tau^3}\left(\textbf{\textit{e}}_{\bf{1}}-3\frac{\overline{x}_1}{\tau^2}(\overline{{\textbf{\textit{x}}}}+2\textbf{\textit{e}}_{\bf{3}})\right)\,\\
\nonumber
 \label{ustressp}
  \overline{\hat{\textbf{\textit{U}}}}_{str\parallel}&=&\left(\frac{\overline{x}_3}{\overline{r}^5}+\frac{2+\overline{x}_3}{\tau^5}\right)\overline{x}_1\overline{\textbf{\textit{x}}}\\
  &+&2\frac{(1+\overline{x}_3)}{\tau^5}\left((2+\overline{x}_3)\textbf{\textit{e}}_{\bf{1}}+\overline{x}_1\textbf{\textit{e}}_{\bf{3}}-5\frac{\overline{x}_1}{\tau^2}(2+\overline{x}_3)(\overline{{\textbf{\textit{x}}}}+2\textbf{\textit{e}}_{\bf{3}})\right)\,.
\end{eqnarray}
The volume integral involved in the computation of the drag correction ${\textbf{\textit{F}}}_{D\alpha}\cdot\textbf{\textit{e}}_{\bf{1}}$ resulting from the transport of the Stokeslet associated with the radial slip velocity by the base straining flow and \textit{vice versa} is similar to that in (\ref{FI1}) with $ \overline{\textbf{\textit{U}}}_{str}$ and $ \overline{\hat{\textbf{\textit{U}}}}_{Sto}$ both replaced by $ \overline{\hat{\textbf{\textit{U}}}}_{Sto\parallel}$. The value provided by numerical integration was $5.5002\times\pi\approx\frac{11}{2}\pi$. Hence at leading order
\begin{equation}
{\textbf{\textit{F}}}_{D\alpha0}\cdot\textbf{\textit{e}}_{\bf{1}}=-\frac{99}{32}\pi\alpha_b Re\kappa^{-1}\textbf{\textit{V}}_{S0}\cdot\textbf{\textit{e}}_{\bf{1}}\,.
\label{FDA0p}
\end{equation}
The $\mathcal{O}(\kappa^0)$-corrections to this estimate arise from the first reflection of the Stokeslet, which induces a $1+\frac{9}{8}\kappa$-multiplicative factor in the right-hand side of (\ref{FDA0p}), and from the `inner' terms which provide an additional $-\frac{62}{15}\pi\alpha Re\textbf{\textit{V}}_{S0}\cdot\textbf{\textit{e}}_{\bf{1}}$ contribution. Collecting all terms, (\ref{FDAp}) is obtained at $\mathcal{O}(\kappa^0)$.\\
\indent  The formal expression for the leading-order force resulting from the advection of the axisymmetric stresslet by the shear flow component and \textit{vice versa} is
 \begin{equation}
 {\textbf{\textit{F}}}_{I\delta0}\cdot\textbf{\textit{e}}_{\bf{1}}=\frac{15}{8}Re\alpha_b\alpha_c \rho_0\int_{\overline{\mathcal{V}}}\overline{\hat{\textbf{\textit{U}}}}_{Sto\parallel}\cdot\left\{\overline{\textbf{\textit{U}}}_{str}\cdot\textbf{\textit{e}}_{\bf{3}}\textbf{\textit{e}}_{\bf{1}}+\overline{x}_3\textbf{\textit{e}}_{\bf{1}}\cdot\overline{\nabla}\overline{\textbf{\textit{U}}}_{str}\right\}d\overline{\mathcal{V}}\,.
 \label{FI0p}
 \end{equation}
 The numerical value of the volume integral was found to be $0.2500\times\pi$, from which we inferred that its exact value is $\pi/4$. Taking into account the $1+\frac{9}{16}\kappa$ multiplicative factor resulting from the reflection of the Stokeslet $\overline{\hat{\textbf{\textit{U}}}}_{Sto\parallel}$ then yields (\ref{FIp}).\\
\indent  Finally, the formal expression for the leading-order force due to time variations of the shear flow component `felt' by the particle as it moves is similar to (\ref{FI20}) with $ \overline{\textbf{\textit{U}}}_{str}$ (resp. $ \overline{\hat{\textbf{\textit{U}}}}_{Sto}$) replaced by $\overline{\textbf{\textit{U}}}_{str\parallel}$ (resp. $ \overline{\hat{\textbf{\textit{U}}}}_{Sto\parallel}$). However the pre-factor now results from the evolution of the particle position along both the normal and radial directions.  Hence this pre-factor is now $\frac{15}{8}\frac{d}{dt}(\rho_0(t)\alpha_c(t))=\frac{15}{4}\kappa\frac{\Lambda^2}{(1+\Lambda)^3}(\textbf{\textit{V}}_{S0}\cdot\textbf{\textit{e}}_{\bf{1}}-3\rho_0\frac{\kappa}{1+\Lambda}\textbf{\textit{V}}_{S0}\cdot\textbf{\textit{e}}_{\bf{3}}+\alpha\rho_0\frac{7+2\Lambda}{(1+\Lambda)^2})$, where we have used the fact that $d\rho_0/dt=\alpha^{-1}\textbf{\textit{V}}\cdot\textbf{\textit{e}}_{\bf{1}}$.
The numerical value of the volume integral was found to be $2.2001\times\pi$, \textit{i.e.} virtually $\frac{11}{5}\pi$. Close to the particle, the integrand is odd with respect to $x_3$ but the reflection of the Stokeslet introduces a $1+\frac{9}{16}\kappa$-multiplicative factor. Truncating the result according to the criteria introduced in \S\,\ref{unste} finally yields (\ref{FUp}).\vspace{3mm}\\
\textbf{Declaration of Interests}\\
 The authors report no conflict of interest.
\section*{Acknowledgements}
 \noindent We thank Dr. Qing Li for providing the numerical data used in \S\S\,\ref{compaDNS} and \ref{compan}. Computational resources were provided by the computing meso-centre CALMIP under project \#P1002. 

\bibliographystyle{jfm}
\bibliography{Biblio}                                                                       

\end{document}